\documentclass[aps,prc,onecolumn,superscriptaddress,nofootinbib,tightenlines,
preprintnumbers,showkeys,floatfix]{revtex4-2}

\usepackage{hyperref}
\usepackage{multirow}
\usepackage{amsmath}
\usepackage{siunitx}
\usepackage{xcolor}
\usepackage{wasysym}
\usepackage{placeins}
\usepackage{bm}
\usepackage{graphics,epsfig,graphicx}
\usepackage{amsbsy}
\usepackage{amsfonts}
\usepackage{cancel}
\usepackage{braket}
\usepackage[T1]{fontenc}
\usepackage{stackengine}
\usepackage{isotope}
\usepackage{environ}
\usepackage{array}
\usepackage{booktabs}
\usepackage{pgfplotstable}

\def\bmr{{\bm r}}
\def\bmx{{\bm x}}
\def\bmy{{\bm y}}
\def\bmz{{\bm z}}

\newcommand{\MeVc}{\ensuremath{\mathrm{MeV}\kern-0.05em/\kern-0.02em \textit{c}}~}
\newcommand{\GeVc}{\ensuremath{\mathrm{GeV}\kern-0.05em/\kern-0.02em \textit{c}}~}
\newcommand{\GeVcSq}{\ensuremath{\mathrm{GeV}\kern-0.05em/\kern-0.02em \textit{c}^2}~}
\newcommand{\MeVcSq}{\ensuremath{\mathrm{MeV}\kern-0.05em/\kern-0.02em \textit{c}^2}~}

\newcommand{\ie}{\textit{i.e.}}
\newcommand{\eg}{\textit{e.g.}}

\newcommand{\MeV}{\ensuremath{\mathrm{MeV}}}

\newcommand{\MN}{M_N}

\newcommand{\Mpi}{M_\pi}

\newcommand{\dd}{\mathrm{d}}
\newcommand{\ii}{\ensuremath{i}}

\newcommand{\vD}{\boldsymbol{D}}
\newcommand{\hc}{\mathrm{h.c.}}

\newcommand{\vq}{\mathbf{q}}

\newcommand{\veck}{\mathbf{k}}

\newcommand{\vecu}{\mathbf{u}}

\newcommand{\couple}[3]{\left({#1}{#2}\right)\!{#3}}

\NewEnviron{subalign}[1][]{%
\begin{subequations}\begin{align}
  \BODY
\end{align}\label{#1}\end{subequations}
}

\NewEnviron{spliteq}{%
\begin{equation}\begin{split}
  \BODY
\end{split}\end{equation}
}

\newcommand{\abs}[1]{\left|#1\right|}

\newcommand{\TopRule}{\toprule[1.25pt]}
\newcommand{\BottomRule}{\TopRule}
\newcommand{\MidRule}{\midrule[0.5pt]}
\newcommand{\MidDoubleRule}{\hline\hline}

\renewcommand{\vec}[1]{\mathbf{#1}}

\begin{document}

\title{Role of three-body dynamics in nucleon-deuteron correlation functions}

\author{M.~Viviani}
\email{michele.viviani@pi.infn.it}
\affiliation{Istituto Nazionale di Fisica Nucleare, Largo B. Pontecorvo 3, 56127 Pisa, Italy}
\author{S.~K\"onig}
\email{skoenig@ncsu.edu}
\affiliation{Department of Physics, North Carolina State University,
Raleigh, NC 27695, USA}
\author{A.~Kievsky}
\email{alejandro.kievsky@pi.infn.it}
\affiliation{Istituto Nazionale di Fisica Nucleare, Largo B. Pontecorvo 3, 56127 Pisa, Italy}
\author{L.~E.~Marcucci}
\email{laura.elisa.marcucci@unipi.it}
\affiliation{Dipartimento di Fisica, Universit\`a di Pisa, Largo B. Pontecorvo 3, 56127 Pisa, Italy}
\affiliation{Istituto Nazionale di Fisica Nucleare, Largo B. Pontecorvo 3, 56127 Pisa, Italy}
\author{B.~Singh}
\email{bhawani.singh@tum.de}
\affiliation{Physics Department, TUM, James-Franck-Stra{\ss}e 1, Garching bei M\"{u}nchen, 85748, 
Germany}
\author{O.~V\'azquez Doce}
\email{oton.vazquezdoce@lnf.infn.it}
\affiliation{Istituto Nazionale di Fisica Nucleare,  Laboratori Nazionali di Frascati, Via Enrico Fermi 54, Frascati 00044, Italy}

\begin{abstract}
Correlation functions of hadrons can be accessed in high-energy collisions of atomic
nuclei, revealing information about the underlying interaction.
This work complements experimental efforts to study nucleon-deuteron $Nd$---with
$N=p$ (proton) or $N=n$ (neutron)---correlations with theory evaluations using
different techniques.
The correlation functions $C_{nd}$ and $C_{pd}$ are calculated based on
scattering wave function, extending previous benchmarks for the $Nd$ scattering
matrix to this new observable.
We use hyperspherical harmonics and Faddeev techniques with one of the widely
used nucleon-nucleon ($NN$) interactions, the Argonne $v_{18}$ potential.
Moreover, in the low-energy region we perform additional calculations in the
framework of pionless effective field theory.
The $pd$ correlation function is computed in the large-energy region to make
contact with a recent measurement by the ALICE Collaboration.
We show that the scattering wave function has the proper dynamical input to describe
an initial rise and subsequent oscillations of $C_{pd}$ as a function of the
energy.
Effects on the observables using different $NN$ and three-nucleon potentials are
evaluated with the conclusion that variations of around $2\%$ are observed.
Although these effects are small, future measurements can go beyond this accuracy
allowing for new detailed studies of strong interaction in light nuclear systems.
The present study supports the current efforts devoted to the measurement of correlation
functions in systems dominated by the strong interactions, such as $pd$,
$ppp$, $\Lambda d$ and $pp\Lambda$.
\end{abstract}

\maketitle

\section{Introduction}

The study and description of the dynamics of few-body systems comprised of nucleons and
atomic nuclei play a fundamental role in nuclear physics.
A comprehensive understanding of the nuclear force between nucleons inside nuclei and in a
denser environment requires an accurate understanding of the few-body dynamics, and achieving
this has been a long-standing goal of nuclear physics for many years.
In this regard, few- and many-body systems provide a unique laboratory for studying nuclear
interactions and the equation of the state of dense nuclear matter~\cite{Bombaci2018,Drischler2021,Lattimer2021}.
In recent decades, modern many-body nuclear physics advances have opened doors to study
nucleon-nucleus and nucleus-nucleus scattering and reactions from first
principles (see Refs.~\cite{Deltuva2008,Leidemann2013,Navratil:2016ycn,Navratil:2022lvq} and references therein).
Nucleon-deuteron ($Nd$) scattering is a fundamental process in nuclear physics that plays a
crucial role in understanding structure and dynamics of atomic nuclei.
The deuteron, composed of one proton and one neutron, is the simplest and most abundant
nucleus in nature that is bound together by a strong force.
Studying its interaction with another nucleon serves as a test bed for studying
three-nucleon effects.
This includes not only the occurrence of genuine three-nucleon forces (3NFs)---\ie, effects
arising not merely from pairwise two-body interactions---but also the interplay of the
nuclear interaction with the Pauli principle, \ie, the fact that nucleons as Fermions need to
have fully antisymmetric wave functions at the quantum level.
The three-nucleon system is the simplest system where these effects can be studied,
and several methods exist to investigate it theoretically with great precision.

Experimentally, the study of the $Nd$ process has a long history dating back to the early 
days of nuclear physics, which includes various techniques, such as scattering experiments, 
polarization measurements, and reactions induced by electroweak probes~\cite{glocke1996,golak2005}.
These measurements have revealed a rich and complex structure that is determined by the
interplay between various aspects of the nuclear force, such as one-pion exchange,
repulsive and attractive components, and the tensor force.
Moreover, in the $pd$ case, the Coulomb force has to be considered as well as, and its effect
is particularly relevant at low energies~\cite{Wood2002}.

Theoretically, $Nd$ scattering was initially studied based on potential models.
In the 1990s, realistic nuclear potentials describing the $NN$ interactions were constructed by
fitting the $NN$ world database~\cite{Av18,Nijmegen,CDBonn}, consisting of more than 4000 $pp$ and
$np$ data at the time, with a $\chi^2$ per datum close to one.
With the development of the effective field theories (EFTs), potential models appeared
constructed order by order in a series expansion of the interaction.
At present, EFT potentials based on chiral effective field theory (Chiral EFT) are available going up to fifth
order in the expansion and they reproduce the $NN$ scattering data up to 350 MeV
lab energy with extremely good accuracy~\cite{Reinert:2017usi,Saha:2022oep}.
Starting with Ref.~\cite{Nogga:2005hy}, focus in the development of
Chiral EFT has shifted towards applying the expansion
directly to observables, investigating in particular questions of
renormalization and details of the so-called ``power counting,'' \ie, the process of assigning orders to
individual components of the interaction.
For a review of these efforts and of nuclear effective field theories in general, see Ref.~\cite{Hammer:2019poc}.

Beyond the two-nucleon sector that is for the most part used to constrain parameters
of the interaction, the three-nucleon system is the simplest scenario in which highly
accurate nuclear potentials can be used to make predictions.
For this reason strong efforts have been made in the last years to solve the equations that
govern the three-nucleon dynamics in bound states and scattering processes.
The necessity of including 3NFs was known for a long time,  mainly motivated by the
fact that highly accurate two-nucleon potentials would typically predict the 
triton binding energy below its experimentally known value.
One of the early successes of nuclear EFTs was the \textit{a priori} prediction that
such forces are present and in fact required.
The impact of the 3NFs in the three- and four-nucleon continuum is at present under
investigation.
In particular, although the chiral expansion nicely organizes the importance of
different three-nucleon interaction terms, only those terms appearing at the lowest orders
have been considered so far. 
There are indications that subleading three-nucleon 
interaction terms, though small, improve the description of particular 
polarization observables~\cite{girlanda2023}. 

In the present study, we focus on the $Nd$ correlation function (defined in more detail in the next section).
The primary motivation for this work is that precise measurements of correlations in the momentum space for the $pd$ system have been made available by the ALICE Collaboration using proton-proton ($pp$) collisions at the Large Hadron Collider (LHC).
For the simpler case of $pp$ correlations, the ALICE Collaboration has previously measured the correlation function using the so-called femtoscopy technique~\cite{ALICE:2018ysd,ALICE:2019buq,ALICE:2020ibs}.
The experimental data can be compared with theoretical calculations by evaluating the integral
\begin{equation}
 C_{pp}(k) = \int \dd^3 r \, S(\vec{r}) \, \abs{\psi_{k}(\vec{r})}^2 \ ,
\label{eq:Cf-pp}
\end{equation}
where $S(\vec{r})$ is a source function that parameterizes
the distance $\vec{r}$ at which the two protons are emitted after the high-energy 
collision, and $\psi_{k}$ is the $pp$ scattering wave function
depending on the reduced relative momentum of the pair 
($k = \abs{\vec{p}_2-\vec{p}_1}/2$).
An extremely accurate description of this observable was obtained using the Argonne
$v_{18}$ (AV18) $NN$ interaction~\cite{Av18}.
As we discuss in detail in the following section, the extension of the formalism to handle the
$Nd$ case is not trivial because the wave function needs to account for the three-nucleon
dynamics mentioned above.
We develop in this work a framework that includes all relevant complexities related to the
correct description of the system, including antisymmetrization effects.
Moreover, the concept of the source function, which for the  $pp$ correlation function is
related to the emission of two nucleons and can be precisely characterized~\cite{ALICE:2020ibs}, has to be
extended to the case of three emitted nucleons, two of which form a deuteron.

Since this is the first time that this observable is analysed for a three-nucleon system,
we find it useful to employ two different approaches to solve the three-body dynamics, the
solution of the Faddeev equations and the hyperspherical harmonic (HH) technique.
Moreover, for the nuclear interaction we consider both the AV18 potential, supplemented by the Urbana IX (UIX) 3NF~\cite{UrbanaIX},
and, to make contact with the modern EFT description of the nuclear interaction, we additionally
use pionless effective field theory (Pionless EFT) to calculate the correlation function.
This EFT has the advantage that its power counting and renormalization
properties are well understood.
However, since by construction this theory does not explicitly include
the physics of pion exchange, its regime of validity it limited to the
low-energy region.
In addition,  we also perform a calculation using an two- and three-nucleon potential model derived
within Chiral EFT, one of the so-called Norfolk interactions~\cite{Norfolk}, the NVIa+3N in the notation of Ref.~\cite{Baroni2018}
In the analysis of the correlation function with such highly accurate $NN$ potentials,
particular attention will be given to the effects of the accompanying three-nucleon forces.

The manuscript is organized in the following way: in the next section, we describe the
theoretical formalism for the calculation of the of $Nd$ correlation functions.
In particular, in Sec.~\ref{sec:Mrow} we discuss the basic formalism for three-nucleon
correlations before we move on to review the HH (Sec.~\ref{sec:cf_pisa_model}) and
Faddeev (Sec.~\ref{sec:PionlessFaddeev}) formalisms, along with a brief introduction
to Pionless EFT.
The main results and benchmarks comparing the different techniques used are given in
Sec.~\ref{sec:Results}.
We conclude with a summary and outlook in Sec.~\ref{sec:Summary}. 
\section{Formalism}

\subsection{Full three-body calculations of the nucleon-deuteron correlation function}
\label{sec:Mrow}

The two-particle femtoscopic correlation function is defined as the ratio of the
Lorentz-invariant yield of a particle pair to the product of the single-particle
yields.
Using $\vec{p}_i$ to denote the momentum of each particle, it can be written as~\cite{Heinz:1999rw}.
\begin{equation}
 C\left(\vec{p}_1, \vec{p}_2\right)
 = \frac{E_1 E_2\, \dd N^{12} /\left(\dd^3 p_1\,\dd^3 p_2\right)}
 {\left(E_1\,\dd N^1 / \dd^3 p_1\right) \left(E_2\,\dd N^2 / \dd^3 p_2\right)}
 = \frac{\mathcal{P}\left(\vec{p}_1, \vec{p}_2\right)}
 {\mathcal{P}\left(\vec{p}_1\right) \mathcal {P}\left(\vec{p}_2\right)} \ .
\label{eq:Ck_basic}
\end{equation}
As indicated by the final equality in Eq.~\eqref{eq:Ck_basic}, the correlation
function can also be understood as the ratio between $\mathcal{P}(\vec{p}_1, \vec{p}_2)$,
the probability of finding a pair of particles with momenta $\vec{p}_1$ and $\vec{p}_2$,
and $\mathcal{P}(\vec{p}_i)$, the probability of finding each particle with
momentum $\vec{p}_i$.
In the absence of any correlations, the two-particle probability factorizes,
$\mathcal{P}(\vec{p}_1, \vec{p}_2) = \mathcal{P}(\vec{p}_1) \mathcal{P}(\vec{p}_2)$,
and the correlation function is equal to unity.
In the quantum mechanical description, the correlation between a pair of particles
(with spins $s_1$ and $s_2$, respectively) can be related to the particle emission
and the subsequent interaction of the particle pair, as discussed in
Ref.~\cite{Mrowczynski:2020ugu}, as
\begin{equation}
 C\left(\vec{p}_1, \vec{p}_2\right)
 = \frac{1}{\Gamma} \sum_{m_1,m_2} \int \dd^3 r_1\,\dd^3 r_2 S_1
 \left(r_1\right) S_1\left(r_2\right)
 \abs{\Psi_{m_1,m_2}(\vec{p}_1,\vec{p}_2,\vec{r}_1,\vec{r}_2)}^2\ ,
\label{eq:2pp}
\end{equation}
where $\Psi_{m_1,m_2}(\vec{p}_1,\vec{p}_2,\vec{r}_1,\vec{r}_2)$ denotes the
two-particle scattering wave function that asymptotically describes 
particle 1 (2) with momentum $\vec{p}_1$ ($\vec{p}_2$) and
spin projection $m_1$ ($m_2$), with weights $\Gamma=(2s_1+1)(2s_2+1)$.
In Eq.~\eqref{eq:2pp} $S_1(r)$ describes the spatial shape of the source for
single-particle emissions.
It can be approximated as a Gaussian probability distribution with a width
$R_M$, which is defined as follows:
\begin{equation}
  S_1(r) = \frac{1}{(2\pi R_M^2)^{\frac32}} e^{{-}r^2/2 R_M^2}\ ,
\label{eq:mrowSignleParicleSource}    
\end{equation}
$R_M$ is also known as the source size for single particle emission.
Eq.~\eqref{eq:2pp} can be simplified by noting that in the wave
functions the dependence on the overall center-of-mass (CM) coordinate
can be trivially factored out.
Introducing the CM coordinate $\vec{R} \equiv \frac{M_1 \vec{r}_1+M_2
\vec{r}_2}{M_1+M_2}$, where $M_1$ and $M_2$ are the masses of the two
particles, the relative distance $\vec{r} \equiv \vec{r}_1-\vec{r}_2$,
and rewriting the two-particle wave function as
$\Psi_{m_1,m_2}(\vec{p}_1,\vec{p}_2,\vec{r}_1,\vec{r}_2)
= e^{{-}i\vec{R}\cdot\vec{P}}\psi_{m_1,m_2,\vec{k}}(\vec{r})$ leads to
the Koonin-Pratt relation for two-particle correlation function~\cite{KOONIN197743},
which we write here as
\begin{equation}
 C(k) = \frac{1}{\Gamma}\sum_{m_1,m_2}\int \dd^3 r \, S(r)
 \abs{\psi_{m_1,m_2,\vec{k}}\left(\vec{r}\right)}^2 \ ,
\end{equation}
where $\psi_{m_1,m_2,\vec{k}}\left(\vec{r}\right)$ represents the
two-particle relative wave function, with $\vec{k}=(\vec{p}_1-\vec{p}_2)/2$, and $S(r)$ is the two-particle emission source, given by
\begin{equation}
 S(r) = \left(\frac{1}{4 \pi R_{\mathrm{M}}^2}\right)^{\!3/2} 
 e^{{-}\frac{r^2}{4 R_{\mathrm{M}}^2}} \ .
\label{eq:twoParticleSource}
\end{equation}
Overall, we have arrived, essentially, at Eq.~\eqref{eq:Cf-pp} as stated
in the Introduction.
We note that for simplicity we did not consider spin degrees of freedom
in writing Eq.~\eqref{eq:Cf-pp}, and moreover the scattering wave 
function $\psi_k$ used in the introduction includes all partial waves.
A more detailed discussion of how the partial-wave expanded form can be
obtained from this will be given for the three-body case below.

For the extension of the formalism to calculate three-nucleon correlation
functions, we follow the general coalescence model as it has been discussed in Ref.~\cite{Mrowczynski:2020ugu}.
For the specific case of nucleon-deuteron correlations, the formalism is
based on the following expressions:
\begin{subalign}[eqs:mrow]
 A_d C_{Nd}(k) &= \frac{1}{6} \sum_{m_2, m_1}
 \int \dd^3r_1 \dd^3r_2 \dd^3r_3\; S_1(r_1) S_1(r_2) S_1(r_3)
 \abs{\Psi^{Nd}_{m_2,m_1,\vec{k}}}^2 \ ,
\label{eq:mrow1} \\
 A_d &= \frac{1}{3} \sum_{m_2}
 \int \dd^3r_1 \dd^3r_2 \; S_1(r_1) S_1(r_2) \abs{\varphi^d_{m_2}}^2\ ,
 \label{eq:mrow2} 
\end{subalign}
where $\vec{k}$ is the $Nd$ relative momentum and  $A_d$ the ``probability of formation of the deuteron''.
The subscript $N$ represents either a proton ($p$) or a neutron ($n$), and
in the following we consider both cases. Hereafter $m_1$ ($m_2$) denotes the spin projection of the nucleon (deuteron). 
We also denote the deuteron bound-state wave function as $\varphi^{d}_{m_2}$, whereas
$\Psi^{Nd}_{m_2,m_1,\vec{k}}$ represents the nucleon-deuteron scattering wave function.
In the above equations we indicate explicitly sums over angular-momentum 
components $m_1$ and $m_2$, but we note that whether or
not these appear explicitly in practical calculations depends on the method
used to solve the equations: we consider both an approach based on HH, where $m_1$ and $m_2$ are explicitly summed over, as well as
Faddeev equations in momentum space, where these sums are implicit in the
choice of basis.

Eq.~\eqref{eqs:mrow} can be simplified by introducing the CM and relative coordinate, as well.
For $A_d$, we change integration variables, introducing 
$\bm{r}=\bm{r}_1-\bm{r}_2$ and $\bm{R}={\frac12}(\bm{r}_1+\bm{r}_2)$ (we disregard the proton-neutron mass difference in this paper).
Writing the product $S_1(r_1) S_1(r_2)$ in terms of $r$ and $R$ and then integrating over $\bm{R}$, one obtains
\begin{equation}
 A_d = \frac{1}{3} \sum_{m_2}
 \int \dd^3r \; \frac{e^{{-}r^2/4R_M^2}}{(4\pi R_M^2)^{\frac32}} 
 \abs{\varphi^d_{m_2}}^2\ .
\label{eq:mrow4}
\end{equation}
In the integral~(\ref{eq:mrow1}), we can use the variables
\begin{equation}
 \bm{x}=\bm{r}_1-\bm{r}_2\ ,
 \qquad \bm{y}=\bm{r}_3-\frac{\bm{r}_1+\bm{r}_2}{2}\ ,
 \qquad \bm{R}_3=\frac13(\bm{r}_1+\bm{r}_2+\bm{r}_3)\ .
\end{equation}
Now
\begin{equation}
 \dd^3r_1 \dd^3r_2 \dd^3r_3 = \dd^3R_3 \dd^3x \dd^3y\ ,
\end{equation}
and
\begin{equation}
 S_1(r_1) S_1(r_2) S_1(r_3)
 = \frac{e^{-(3R_3^2+{\frac23}y^2+{\frac12}x^2)/2R_M^2}}%
   {(2\pi R_M^2)^{\frac92}} \ .
\end{equation}
Integrating over $d^3R_3$ (the wave function $\Psi^{Nd}_{m_2,m_1,\vec{k}}$ does not depend on $R_3$), we obtain
\begin{equation}
 A_d C_{Nd}(k) = {\frac16}\sum_{m_2, m_1}
 \int \dd^3x\dd^3y\; \frac{e^{-({\frac43}y^2+x^2)/4R_M^2}}%
 {(3\pi R_M^2)^{\frac32}(4\pi R_M^2)^{\frac32}}
 \abs{\Psi^{Nd}_{m_2,m_1,\vec{k}}}^2 \ .
\label{eq:mrow7}
\end{equation}
Introducing the vectors $\bm{\xi}_1=\sqrt{\frac43}\bm{y}$
and $\bm{\xi}_2=\bm{x}$, this integral can be rewritten as
\begin{equation}
 A_d C_{Nd}(k) =  {\frac16} \sum_{m_2, m_1} 
 \left({\frac34}\right)^{\frac32}
 \int \dd^3\xi_1\dd^3\xi_2\; 
 \frac{e^{{-}(\xi_1^2+\xi_2^2)/4R_M^2}}{(3\pi R_M^2)^{\frac32} (4\pi R_M^2)^{\frac32}}
 \abs{\Psi^{Nd}_{m_2,m_1,\vec{k}}}^2  \ . 
\label{eq:mrow8}
\end{equation}
Let us now introduce the so-called hyperradius, defined as $\rho=\sqrt{\xi_1^2+\xi_2^2}$,
and the hyperangles variables $\Omega$~\cite{Marcucci:2019hml} (see below), such that 
$\dd^3\xi_1\dd^3\xi_2=\rho^5 \dd\rho \dd\Omega$.
Finally, we obtain
\begin{equation}
 A_d C_{Nd}(k) =  {\frac16} \sum_{m_2, m_1} \int \rho^5 \dd\rho \dd\Omega\; 
 \frac{e^{{-}\rho^2/4R_M^2}}{(4\pi R_M^2)^3}
 \abs{\Psi^{Nd}_{m_2,m_1\vec{k}}}^2 \ .
\label{eq:mrow9}
\end{equation}
As a check of this formula, let us approximate the $Nd$ wave function by the following asymptotic structure properly anti-symmetrized
\begin{equation}
 \Psi^{Nd}_{m_2,m_1,\vec{k}}= \frac1{\sqrt{3}}\bigl[ 
 e^{i\bm{k}\cdot\bm{y}_3}\varphi^d_{m_2}(1,2) \chi_{m_1}(3)+
 e^{i\bm{k}\cdot\bm{y}_1} \varphi^d_{m_2}(2,3) \chi_{m_1}(1)+
 e^{i\bm{k}\cdot\bm{y}_2} \varphi^d_{m_2}(3,1) \chi_{m_1}(2) ]\ ,
\end{equation}
where $\bm{y}_\ell$ are defined below in Eq.~\eqref{eq:jacobi} (see also Fig.~\ref{fig:jacobi}, note that $\vec{x}_3\equiv\vec{x}$ and  $\vec{y}_3\equiv\vec{y}$).
We work here with the form given in Eq.~\eqref{eq:mrow7}.
For $k\rightarrow\infty$, we can disregard the terms coming from different
permutations in $\abs{\Psi^{Nd}_{m_2,m_1,\vec{k}}}^2$, as
$e^{-i\bm{k}\cdot\bm{y}_3}\times e^{i\bm{k}\cdot\bm{y}_2}$,
as their contribution becomes vanishing.
We have three terms left. However, it results
\begin{equation}
 {\frac43}y^2+x^2= \rho^2 = {\frac43}y_\ell^2+x_\ell^2\, \qquad
 \ell = 1,2,3\ .
\end{equation}
Therefore, each of the three terms gives the same contribution cancelling the factor
$(1/\sqrt{3})^2$, and one obtains
\begin{spliteq}
 A_d C_{Nd}(k) &=  {\frac16} \sum_{m_2, m_1} \int \dd^3x\dd^3y\; 
 \frac{e^{{-}({\frac43}y^2+x^2)/4R_M^2}}%
 {(3\pi R_M^2)^{\frac32} (4\pi R_M^2)^{\frac32}} 
 \abs{\varphi^d_{m_2}(1,2)}^2 \ , \\
 &= {\frac12} \sum_{m_1} \int \dd^3y\;
 \frac{e^{{-}({\frac43}y^2)/4R_M^2}}{(3\pi R_M^2)^{\frac32}} A_d\ , \\
 &= A_d \ .
\end{spliteq}
Therefore, in this case $C_{Nd}(k\rightarrow\infty)=1 $ as expected.
\subsection{Hyperspherical harmonics description of the proton-deuteron wave function }
\label{sec:cf_pisa_model}

\begin{figure}[htbp]
 \centering
  \includegraphics[scale=0.3]{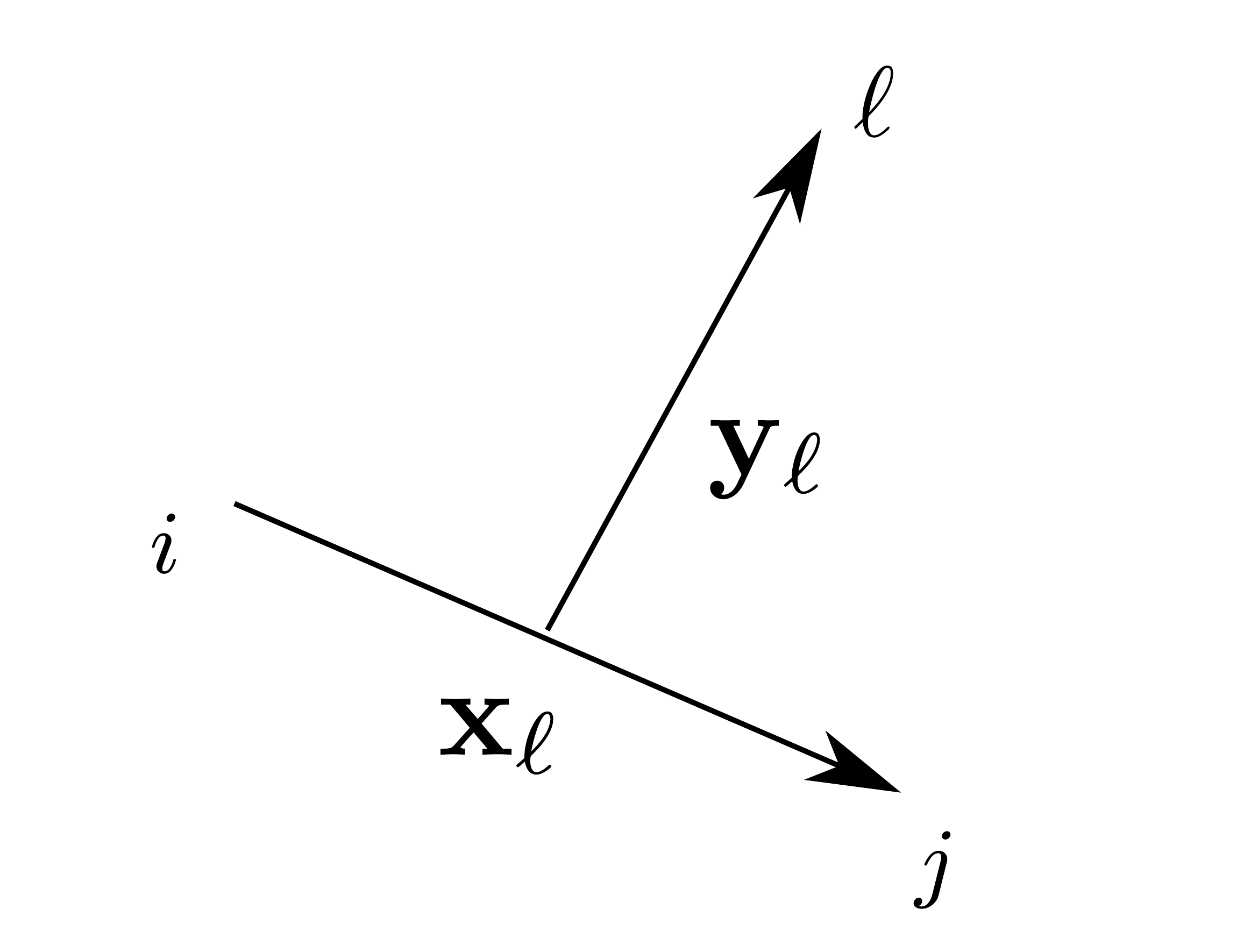}
  \caption{Definition of the Jacobi vector $\bmx_\ell$ and $\bmy_\ell$.}
  \label{fig:jacobi}
\end{figure}

We discuss the calculation of the proton-deuteron correlation function by taking into account the full dynamics of the three particles. Initially, we 
include only the effect of the long-range Coulomb interaction between the proton and the deuteron. However, we still take
into account the antisymmetrization of the wave function. This simple scenario will allow us to subsequently address the full case including the short-range nuclear interaction.
In the following, for a given permutation $ij\ell$ of the three nucleons, we define the Jacobi coordinates (see Fig.~\ref{fig:jacobi})
\begin{equation}
  \bmx_\ell=\bmr_j-\bmr_i\ , \qquad \bmy_\ell=\bmr_\ell-\frac{\bmr_i+\bmr_j}2\ .
\label{eq:jacobi}
\end{equation}
To construct an antisymmetric wave function it is sufficient to consider the three ``even'' permutations of the three particles, namely $ij\ell=123$, $231$ and $312$. For this reason, we can specify the permutation (and the various quantities) just giving the value of $\ell=1,2,3$. 
   
We introduce here also the hyperangular coordinates. The hyperradius $\rho$ and hyperangles $\varphi_\ell$ are defined in terms of the moduli of the Jacobi vectors, explicitly
\begin{equation}
  \rho=\sqrt{x_\ell^2+{\frac43} y_\ell^2}\ , \qquad
   \tan\varphi_\ell=\sqrt{\frac43}{\frac{y_\ell}{x_\ell}}\ .\label{eq:rho}
\end{equation}
The hyperradius $\rho$ turns out to be independent on the permutation $\ell$. In fact, it can be shown that $\rho^2 ={\frac23} (r_{12}^2+r_{13}^2+r_{23}^2)$, where $r_{ij}$ is the distance between particles $i$ and $j$. The set of hyperangular and angular variables
is denoted with $\Omega_\ell$, namely
\begin{equation}
   \Omega_\ell\equiv\{ \varphi_\ell,\hat \bmx_\ell,\hat \bmy_\ell\} \ ,\label{eq:Omega}
   \end{equation}
   where $\hat \bmx_\ell$ ($\hat \bmy_\ell$) denotes the polar angles of vector 
   $\bmx_\ell$ ($\bmy_\ell$). When the permutation index is not indicated
   the reference order of the particles $ij\ell=123$ is understood.

\subsubsection{Free case}

Let us consider the free case, \ie, no nuclear interaction between the $p$ and $d$ clusters. The wave function is then simply
given by 
\begin{equation}
  \Psi_{m_2,m_1,\vec{k}}^{pd,(\text{free})} = \frac{1}{\sqrt{3}} \sum_{\ell}^{\text{even\ perm.}} \varphi^d_{m_2}(i,j) \chi_{m_1}(\ell) \Phi_c({\bm k},{\bm y_\ell})\ ,
  \label{eq:free}
  \end{equation}
where as before $\varphi^d_{m_2}(i,j)$ is the deuteron wave function with spin projection $m_2$, ${\bm k}$ is the relative momentum between the two clusters, $\chi_{m_1}(\ell)$ is a spinor describing the proton,
and $ \Phi_c({\bm k},{\bm y_\ell})$ is a Coulomb-distorted plane wave, having the following partial-wave expansion:
\begin{equation}
  \Phi_c({\bm k},{\bm y}) = \sum_{LM} 4\pi\, i^L\, Y_{LM}^*(\hat {\bm k}) Y_{LM}(\hat \bmy) e^{i\sigma_L} \frac{F_L(\eta,ky)}{ky}\ .
  \label{eq:coulpw}
\end{equation}
Here $F_L(\eta,ky)$ is the regular Coulomb function and $\sigma_L$ the Coulomb phase shift. The neutron-deuteron case can be recovered just replacing $\Phi_c({\bm k},{\bm y_\ell})\rightarrow e^{i\vec{k}\cdot\vec{y}_\ell}$.



Before we address the full interacting case, it is convenient to expand the free wave function~(\ref{eq:free}) using Eq.~(\ref{eq:coulpw})
and rewrite it as a sum of terms with definite total angular momentum $J$. Using one of the possible choice of the
recoupling order, we obtain
\begin{spliteq}
  \Psi^{pd,(\text{free})}_{m_2,m_1,\vec{k}} &= \sum_{LSJ}\sqrt{4\pi} i^L \sqrt{2L+1} e^{i\sigma_L} (1 m_2 {\frac12} m_1 \mid S J_z)
  (L 0 S J_z \mid J J_z) \\
  &\times \frac{1}{\sqrt{3}} \sum_{\ell}^{\text{even\ perm.}}
  \left\{ Y_L(\hat \bmy_\ell) \Bigl[ \varphi^d_{m_2}(i,j) \chi_{m_1}(\ell)\Bigr]_S\right\}_{JJ_z} \frac{F_L(\eta,ky_\ell)}{k y_\ell}\ ,
\end{spliteq}
where we have assumed ${\hat{\bm k}}//{\hat{\bmz}}$ and therefore $Y_{LM}(\hat {\bm k})=\frac{\sqrt{2L+1}}{\sqrt{4\pi}} \delta_{M,0}$.  Note that $J_z=m_1+m_2$. 

\subsubsection{Fully interacting case}

Now, let us consider the fully interacting case. The total wave function becomes
\begin{equation}
  \Psi^{pd}_{m_2,m_1,\vec{k}} =  \sum_{LSJ}\sqrt{4\pi} i^L \sqrt{2L+1} e^{i\sigma_L} (1 m_2 {\frac12} m_1 \mid  S J_z)
      (L 0 S J_z\mid J J_z) \Psi_{LSJJ_z} \ ,\label{eq:full1}
\end{equation}
where $\Psi_{LSJJ_z}$ are three-body wave functions satisfying $(H-E)\Psi_{LSJJ_z}=0$, with
\begin{equation}
  E={\frac43} \frac{k^2}{\MN}-B_d \ ,
\label{eq:E} 
\end{equation}
$B_d$ denoting the deuteron binding energy and $M_N$ the nucleon mass.
We compute such wave functions as
\begin{spliteq}
    \Psi_{LSJJ_z} &= \sum_{n,\alpha} 
    \frac{u_{n,\alpha}(\rho)}{\rho^{5/2}} 
    {\cal Y}_{n,\alpha}(\Omega) \\
     &+ \frac{1}{\sqrt{3}} \sum_{\ell}^{\text{even\ perm.}}
    \left\{ Y_L(\hat \bmy_\ell) \Bigl[ \varphi^d(i,j) \chi(\ell)\Bigr]_S\right\}_{JJ_z} \frac{F_L(\eta,ky_\ell)}{k y_\ell}\\
     &+  \sum_{L'S'} T^J_{LS,L'S'} \frac{1}{\sqrt{3}} \sum_{\ell}^{\text{even\ perm.}}
    \left\{ Y_{L'}(\hat \bmy_\ell) \Bigl[ \varphi^d(i,j) \chi(\ell)\Bigr]_{S'}\right\}_{JJ_z} \\
    & \times \frac{\overline G_{L'}(\eta,ky_\ell)
    +i F_{L'}(\eta,ky_\ell)}{k y_\ell}\ .
    \label{eq:fullJ}
\end{spliteq}
Here ${\cal Y}_{n,\alpha}(\Omega)$ is a set of completely antisymmetric basis functions constructed in terms of HH functions (which form a complete basis in the $\Omega$ Hilbert space) multiplied by appropriate combinations of spin-isospin states of the three particles. Explicitly,
\begin{spliteq}
 {\cal Y}_{n,\alpha}(\Omega) &= \sum_{\ell}^{\text{even\ perm.}} f_\alpha(x_\ell) 
 {\cal N}_\alpha (\sin\phi_\ell)^{L_y} (\cos\phi_\ell)^{L_x} P^{L_y
 +{\frac12},L_x +{\frac12}}_n(\cos2\phi_\ell)\\
 &\times \Bigl\{ \Bigl[Y_{L_y}(\hat y_\ell) Y_{L_x}(\hat x_\ell) \Bigr]_{\Lambda} 
 \Bigl[ (s_i s_j)_{S_2} s_\ell\Bigr]_\Sigma\Bigr\}_{JJ_z} 
 \Bigl[ (t_i t_j)_{T_2} t_\ell\Bigr]_{T,T_z}\ ,\label{eq:phh}
\end{spliteq}
where $\alpha\equiv\{L_x,L_y,\Lambda,S_2,\Sigma,T_2,T\}$ is a set of quantum numbers specifying the HH functions, $P_n^{a,b}$ a Jacobi polynomial of degree $n$, and ${\cal N}$ suitable normalization factors. Moreover, to accelerate the convergence of the expansion over $n$, we have added the so-called ``correlation factors'' $f_\alpha(x_\ell)$, chosen in order to reproduce the behaviour of the wave function when the particles $i$ and $j$ are close (and thus the distance $x_\ell$ is small). These functions therefore describe the two-body correlations of the wave functions, and clearly they depend on the spin-isospin-angular state of the pair (\ie, by the quantum numbers $\alpha$). The expansion so organized is called Pair-Correlated HH (PHH) expansion. For a discussion of the choice of the functions $f_\alpha$ and a review of the properties of the HH and PHH states, see, for example, Refs.~\cite{Kievsky:2008es,Marcucci:2019hml}. 

The calculation is performed as follows. For a given choice of the quantum number $\alpha$ (called a ``channel''), the functions given in Eq.~(\ref{eq:phh}) with increasing values of $n$ are considered, up to a value $N_\alpha$, chosen so to have a convergence of the quantities of interest, as the phase-shifts. Thanks to the presence of the correlation factor, a very good convergence is obtained already with low values of $N_\alpha$, typically $N_\alpha=5-10$ is sufficient. Then other channels are included in the expansion, etc. Note the the most important channels are those with low values of $L_x$ and $L_y$, so we usually start by including in the expansion channels with $L_x+L_y={\cal L}=0$ or $1$, and then increase the values of ${\cal L}$. Usually it is sufficient to consider channels up to ${\cal L}=5$ or $6$, of a total number ranging from $18$ to $30$, depending on the value of $J^\pi$. At the end the expansion over $n$ and $\alpha$ includes some hundreds terms. 

The expansion in the first row of Eq.~(\ref{eq:fullJ}) describes the system when the three nucleons are close to each other. The hyperradial functions $u_{n,\alpha}(\rho)$ are obtained by solving a set of coupled differential equations, obtained using the Kohn variational principle~\cite{Kievsky:2001fq,Kievsky:2004em}. They go asymptotically to zero for energies below the deuteron breakup threshold, whereas for energies $E\equiv Q^2/m>0$, see Eq.~(\ref{eq:E}), the hyperradial functions
$u_{n,\alpha}(\rho)\rightarrow A_\alpha e^{i Q\rho}$ when $\rho\rightarrow\infty$.

Moreover, in Eq.~(\ref{eq:fullJ})  $ T^J_{LS,L'S'}$ are the T-matrix elements, also obtained from the Kohn variational principle, which describe $pd$ scattering observables. Above we have introduced
  \begin{equation}
    \overline G_{L'}(\eta,ky)=G_{L'}(\eta,ky) \Bigl( 1- e^{-\beta y}\Bigr)^{(2L'+1)}\ , 
  \end{equation}
  $G_{L'}$ being the irregular Coulomb function. With this definition, the regularized irregular Coulomb functions $\overline G_{L'}(\eta,ky)$ are well behaved for all
  values of $y$ and for $y\gg\beta^{-1}$ they reduce to the irregular Coulomb functions. Typically $\beta=0.25$ fm${}^{-1}$ is used as regularization scale.
 In Eq.~(\ref{eq:fullJ}),  $L'S'$ are all possible combinations for the given $J$ and parity $(-1)^L$. 

The asymptotic behaviour of the wave functions $\Psi_{LSJJ_z}$ is chosen so that if we turn off the nuclear interaction they reduce to
\begin{equation}
    \Psi_{LSJJ_z}\rightarrow
    \frac{1}{\sqrt{3}} \sum_{\ell}^{\text{even\ perm.}}
    \left\{ Y_L(\hat \bmy_\ell) \Bigl[ \varphi^d(i,j) \chi(\ell)\Bigr]_S\right\}_{JJ_z} \frac{F_L(\eta,ky_\ell)}{k y_\ell}\ .
    \label{eq:freeJ}
\end{equation}
In fact, in such a case, $u_\alpha=T^J_{LS,L'S'}=0$ and $\Psi^{pd}_{m_2,m_1,\vec{k}}$ reduces to $\Psi^{pd,(\text{free})}_{m_2,m_1,\vec{k}}$. Moreover,
the asymptotic behaviour is such that the part multiplying the T-matrix has a form of an outgoing wave, since
$G_{L'}(\eta,ky_\ell)+i F_{L'}(\eta,ky_\ell)\sim e^{i ky_\ell}$.

In the calculation, we will include the effect of the nuclear interaction up to a given $\overline{ J}$. In fact, for $J>\overline{J}$,
the centrifugal barrier should forbid the three particle to be close, in that case the free wave function given
by Eq.~(\ref{eq:freeJ}) should be appropriate. Accordingly, it is convenient to resum all the terms proportional to $F_L(\eta,k y_\ell)$ in order to reproduce the free wave function. Let us define
\begin{spliteq}
    \widetilde\Psi_{LSJJ_z}
    &= \sum_\alpha \frac{u_\alpha(\rho)}{\rho^{5/2}} 
    {\cal Y}_\alpha(\Omega) \\
     &+  \sum_{L'S'} T^J_{LS,L'S'} \frac{1}{\sqrt{3}} \sum_{\ell}^{\text{even\ perm.}}
    \left\{ Y_{L'}(\hat \bmy_\ell) \Bigl[ \varphi^d(i,j) \chi(\ell)\Bigr]_{S'}\right\}_{JJ_z} \\
    & \times \frac{\overline G_{L'}(\eta,ky_\ell)+i F_{L'}(\eta,ky_\ell)}{k y_\ell}\ ,
      \label{eq:fullJm}
\end{spliteq}
  where namely we have subtracted from the wave function given in Eq.~(\ref{eq:fullJ}) the ``free'' part. Then the total wave
  function $\Psi^{Nd}_{m_2,m_1,\vec{k}}$ can be cast in the form
\begin{equation}
  \Psi^{pd}_{m_2,m_1,\vec{k}} =  \Psi^{pd,(\text{free})}_{m_2,m_1,\vec{k}}
   + \sum_{LSJ}^{J\le \overline{J}} \sqrt{4\pi} i^L \sqrt{2L+1} e^{i\sigma_L} (1 m_2 {\frac12} m_1  \mid  S J_z)
      (L 0 S J_z\mid  J J_z) \widetilde\Psi_{LSJJ_z} \ .
\end{equation}
This is the three-nucleon wave function which asymptotically behaves as a $pd$ (distorted) plane-wave, with the proton (deuteron)
in the spin state $m_1$ ($m_2$). The components $\widetilde\Psi_{LSJJ_z}$ describe configurations where the three particles are close to each other. Varying the value of $\overline J$ it is possible to control the waves where the strong interaction is taken into account. 

Great care must be taken in order to include a sufficient number of
PHH states ${\cal Y}_{n,\alpha}(\Omega)$, in particular it is necessary to include a sufficient number of channels $\alpha$ in Eq.~(\ref{eq:fullJm}).
This part is essential to describe the configurations where the three particles
are close to each other. Partial waves where the convergence of this expansion is more critical are those in which the orbital angular momentum $L$ takes its lowest values, $J^\pi=1/2^+$, $3/2^+$, $1/2^-$, $3/2^-$, and $5/2^-$. For the $J^\pi=1/2^+$ case, there is the formation of the ${}^3$He bound state and therefore the scattering wave function must be constructed orthogonal to it. For the states with relative orbital angular momentum $L=1$, the interaction between the three particles is rather attractive, therefore a large number of terms  in the sum over $\alpha$ in Eq.~(\ref{eq:fullJm}) is required.

\subsection{Pionless EFT with momentum-space integral equations}
\label{sec:PionlessFaddeev}

Pionless effective field theory (EFT) is designed to capture the universal
low-energy features of few-nucleon systems that arise from the fact that the
nucleon-nucleon $S$-wave scattering lengths are large compared to the typical
range of the nuclear interaction, set by the inverse pion mass $M_\pi^{-1} \sim
1.4$~fm.
The theory is constructed to yield the most general parametrization of the
nuclear force within its regime of validity (characterized by the EFT breakdown
scale $\sim M_\pi$), and it has been used to make a number of highly precise
predictions for low-energy processes (for a recent review of the theory and applications, see Ref.~\cite{Hammer:2019poc}).

The strong nuclear interaction in Pionless EFT is described by a series of contact
(zero-range) interactions, including an increasing number of derivatives as one
goes to subsequently higher order in the EFT expansion.
In the two-nucleon sector, this series reproduces the well-known effective
range expansion~\cite{Bethe:1949yr}, which Pionless EFT consistently extends to few- and
many-nucleon systems.
Pionless EFT in particular captures the universal physics reflecting the
closeness of low-energy nuclear systems to the unitarity limit (infinite two-nucleon 
$S$-wave scattering lengths).
A remarkable feature stemming from this is the appearance of a three-nucleon
contact interaction at leading order in the theory~\cite{Bedaque:1998kg,Bedaque:1999ve},
which one would naively expect to be subleading.

In order to evaluate the correlation function as defined in Sec.~\ref{sec:Mrow} in
momentum space, in the following subsection we consider first the general Faddeev formalism for scattering
calculations, which is not limited to interactions derived from Pionless EFT.
Since for the moment we neglect electromagnetic effects, the resulting expressions
will be valid for calculations of neutron-deuteron correlation functions.
For the proton-deuteron system, discussed subsequently in Sec.~\ref{sec:Pionless-pd}, we will instead use equations derived directly from
a diagrammatic approach, which we relate to general formalism.

\subsubsection{Faddeev approach for neutron-deuteron scattering}
\label{sec:Faddeev}

We will follow here largely Ref.~\cite{Huber:1995zza}, with some 
differences, and begin with an overview of the homogeneous Faddeev 
equation that describes bound states.
Neglecting three-nucleon forces, the basic Faddeev equation for the
three-nucleon bound-state problem can be written as
\begin{equation}
 \ket{\psi} = G_0 t P \ket{\psi} \ ,
\label{eq:fadeq}
\end{equation}
where $\ket{\psi}$ denotes one of three equivalent Faddeev components, $G_0$ is
the free three-nucleon Green's function, $t$ denotes the two-nucleon T-matrix,
and $P$ is a permutation operator defined as
\begin{equation}
 P = P_{12}P_{23} + P_{13}P_{23} \ .
\label{eq:P}
\end{equation}
For definiteness, we use here the convention that $\ket{\psi}$ is the Faddeev
components with nucleons 1 and 2 singled out.
Therefore, $t$ acts within the $(12)$ subsystem, and in order to represent the
equations in momentum space we use Jacobi momenta $\vecu_1 =
\frac12(\veck_1-\veck_2)$ and
$\vecu_2 = \frac23[\veck_3-\frac12(\veck_1+\veck_2)]$, where $\veck_i$ are the
individual nucleon momenta.
Note that $\vecu_1$ and $\vecu_2$ are the momenta conjugate to the Jacobi
coordinated $\vec{x}$ and $\vec{y}$, repsectively, that were introduced
previously.
Projecting the three-dimensional momenta onto partial waves yields states $\ket{u_1u_2;s}$,
where
\begin{equation}
 \ket{s} = \ket{%
  \couple{l_2}{\couple{\couple{l_1}{s_1}{j_1}}{\tfrac12}{s_2}}{J};
  \couple{t_1}{\tfrac12}{T}
 }
\label{eq:s}
\end{equation}
collects angular momentum, spin, and isospin quantum numbers.
They are coupled such that $\couple{l_1}{s_1}{j_1}$ and $t_1$ describe the
two-nucleon subsystem, whereas $l_2$ denotes the orbital angular momentum
associated with the Jacobi momentum $u_2$ and $s_2$ is an intermediate quantum
number.
Given a solution $\ket{\psi}$ of Eq.~\eqref{eq:fadeq},
the full three-nucleon wave function can be obtained as
\begin{equation}
 \ket{\Psi} = (1+P) \ket{\psi} \ .
\label{eq:Psi-psi}
\end{equation}

In order to calculate three-nucleon scattering, we need to consider an
inhomogeneous Faddeev equation.
Specifically, we are interested here in neutron-deuteron ($nd$) scattering, and
to set up that system we define a state
\begin{equation}
 \ket{\phi} = \ket{\varphi_d k;s_d}
\label{eq:phi}
\end{equation}
that is a product of a deuteron state $\ket{\varphi_d}$ in the $(12)$ subsystem and
a plane wave $\ket{k}$ which describes the relative motion of the third nucleon
with respect to the deuteron.
The $s_{d}$ in Eq.~\eqref{eq:phi} is used to denote a set of three-nucleon
quantum numbers restricted to channels that support the existence of a deuteron
component, \ie, $\ket{s_{d}}$ necessarily has $s_1 = 1$, $t_1 = 0$, $j_1=1$
and $l_1 = 0$ or $2$.
In the momentum-space partial-wave representation, we have
\begin{equation}
 \braket{u_1 u_2;s|\phi}
 \sim \varphi_d^{(l_1)}(u_1) \frac{\delta(u_2 - k)}{u_2^2} \delta_{s,s_{d}} \ ,
\end{equation}
where $\varphi_d^{(l_1)}(u_1)$ is the momentum-space wave function of the
deuteron component with angular momentum $l_1$.
In configuration space, if $\bmy$ denotes the Jacobi coordinate conjugate to
$\vecu_2$, the representation of $\ket{\phi}$ involves a spherical Bessel
function $j_{l_2}(k y)$.\footnote{%
For $pd$ scattering the Bessel function would be replaced by a regular Coulomb wave function.}

With the help of $\ket{\phi}$ we can now introduce an operator $\tilde{T}$ that
satisfies
\begin{equation}
 \ket{\psi_k;s_{d}} = \ket{\phi} + \tilde{T}\ket{\phi} \ ,
\label{eq:psi-phi-T}
\end{equation}
where $\ket{\psi_k}$ is one Faddeev component of the neutron-deuteron
scattering state with relative momentum $k$.
Note that this definition is analogous to the definition of the two-body
T-matrix as the operator that maps a plane-wave state to the full scattering
state with the same momentum.
The $\tilde{T}$ we use here is related to the operator called $T$ in
Ref.~\cite{Huber:1995zza} by $\tilde{T} = G_0 T$.
The inhomogeneous Faddeev equation used to calculate $\tilde{T}$ has the form
\begin{equation}
 \tilde{T}\ket{\phi} = G_0 t P\ket{\phi} + G_0 t P \tilde{T}\ket{\phi} \ .
\label{eq:fadeq-T-phi}
\end{equation}
For clarity we choose here, unlike most references on the subject, to
explicitly write the dependence on $\ket{\phi}$, so really the object that we
obtain by solving Eq.~\eqref{eq:fadeq-T-phi} is $\tilde{T}\ket{\phi}$.
Note that working with $\tilde{T}$ instead of $T$ is convenient for our goal of
calculating scattering wave functions, but it is by no means a necessary choice:
$T$ and $\tilde{T}$ contain exactly the same physics information and one can
easily be obtained from the other.
For a numerical solution we project Eq.~\eqref{eq:fadeq-T-phi} onto the
momentum-space partial-wave states $\ket{u_1u_2;s}$ introduced before.
To that end, note that only the total spin $J$ and isospin $T$ (and their
projections $M_J$ and $M_T$ that we do not specify explicitly) are conserved
quantum numbers for the three-nucleon system.
Therefore, in practice we need to fix $J$ and $T$ and include \emph{all}
channels $\ket{s}$ for which the intermediate quantum numbers defined in
Eq.~\eqref{eq:s} can couple to the chosen total $J$ and $T$.
From Eq.~\eqref{eq:fadeq-T-phi} one therefore obtains a set of
\emph{coupled} integral equations, which turn into a set of coupled matrix
equations upon discretization of the Jacobi momenta $u_{1,2}$ on a quadrature
mesh.
We omit here the details of that numerical procedure and instead focus on how
to obtain scattering parameters and wave functions from a solution of the
equation system obtained via Eq.~\eqref{eq:fadeq-T-phi}.

In order to obtain elastic scattering parameters, one calculates from
$\tilde{T}$ another quantity
\begin{equation}
 U\ket{\phi} = P G_0^{{-}1} \ket{\phi} + P \tilde{T} G_0^{{-}1} \ket{\phi} \ ,
\label{eq:U}
\end{equation}
and then this needs to be contracted with $\bra{\phi'} = \bra{\varphi_d
k;s_{d}'}$ from the left to obtain matrix elements $\braket{\phi'|U|\phi}$.
The dimension of the final matrix is determined by the allowed combinations of
quantum numbers $l_2$ and $s_2$ for a given fixed total $J$, whereas $l_1$ is
summed over for each individual matrix element.
If standing-wave boundary conditions are chosen for the solution of
Eq.~\eqref{eq:fadeq-T-phi}, the resulting matrix is a $K$ matrix from which it
is straightforward to obtain phase shifts and mixing angles after picking a
particular representation.

The procedure for calculating scattering wave functions is slightly different.
Firstly, Eq.~\eqref{eq:fadeq-T-phi} is most conveniently solved with outgoing
boundary conditions in order to have direct access to the imaginary part of the
amplitude.
Instead of $U$ we are now interested
directly in the Faddeev component $\ket{\psi_k;s_{d}}$ as defined in
Eq.~\eqref{eq:psi-phi-T}.
In order to obtain from this a relative $nd$ wave function in momentum space, we
need to project onto an outgoing asymptotic state similarly to what we did to
obtain $\braket{\phi'|U|\phi}$, except that now we are using
$\bra{\phi'} = \bra{\varphi_d u_2;s_{d}'}$, with an arbitrary momentum $u_2$
and $\bra{s_{d}'}$ such that $l_2' = l_2$ and $s_2' = s_2$.
Assuming that $\ket{\varphi_d}$ is properly normalized to unity, this yields an
expression of the form
\begin{equation}
 \braket{\varphi_d u_2;s_{d}'|\psi_k;s_{nd}}
 = \frac{\delta(u_2 - k)}{k^2}
 + \braket{\varphi_d u_2;s_{d}'|\tilde{T}|\varphi_d k;s_{d}} \ .
\label{eq:psi-T}
\end{equation}

At this point we note that the discussion so far is 
based only on the single Faddeev component $\ket{\psi_k;s_{d}}$ and the
corresponding amplitude $\tilde{T}$.
That is sufficient if one is interested merely in extracting elastic scattering
information (via $U$), and a Fourier-Bessel transformation of
Eq.~\eqref{eq:psi-T} will produce a wave function the (reduced) radial part of
which has the appropriate form $\sim \sin(ky+\delta(k))$, which is used for
example in configuration-space formulations of the Faddeev
equations~\cite{Chen:1989zzc}.
To actually calculate the \emph{full} scattering wave function, however, we need
to use the analog of Eq.~\eqref{eq:Psi-psi} for scattering calculations, \ie,
\begin{equation}
 \ket{\Psi_k;s_{d}} = (1 + P) \ket{\psi_k;s_{d}} \ .
\label{eq:Psi-k-sd}
\end{equation}
Based on this we can then proceed as before and project onto $\bra{\phi'} =
\bra{\varphi_d u_2;s_{d}'}$.
The result involves the same distribution part $\delta(u_2-k)/k^2$ (which can
be seen directly), and its Fourier-Bessel transform will exhibit the same
asymptotic behavior $\sim \sin(ky+\delta(k))$, but the antisymmetrization
changes the detailed structure at short distances.

In order to evaluate the correlation function within this formulation (without
Fourier transformation of the wave functions to coordinate space), we note that the
source function $S(r)$ can be written as an operator $\hat{S}$ that is local
in coordinate space:
\begin{equation}
\bra{\vec{r}}\hat{S}\ket{\vec{r}'} = \frac{\exp\!\big({-}r^2/R^2\big)}{(4\pi R)^{3/2}} \delta^{(3)}(\vec{r}-\vec{r}') \ .
\end{equation}
In momentum space, this translates to a non-local representation that can be written
in closed form~\cite{Tabakin:1966aa}:
\begin{equation}
{\bra{q,\ell}\hat{S}\ket{q',\ell'} = \exp\left({-}R^2(q^2+q'^2)\right)\,i_\ell\left(2Rqq'\right)\delta_{\ell\ell'}}
\end{equation}
In this expression, $\ell$ denotes the orbital angular momentum of a particular 
partial wave, and $i_\ell$ is a modified spherical Bessel function.
The scale $R$ is related to the source radius of Sec.~\ref{sec:Mrow}
via $R = \sqrt{3/4}R_M$.
Overall, we can now write Eqs.~\eqref{eq:mrow1} and~\eqref{eq:mrow2} as
\begin{subalign}[eqs:mrow-op]
 A_d \, C_{nd}(k) &= \sum_{s_{d}} \alpha(s_{d}) \bra{\Psi_k;s_{d}}{\hat{S}}\ket{\Psi_k;s_{d}} \ ,
 \label{eq:mrow-op-Cnd} \\
 A_d &= \bra{\varphi_d}{\hat{S}}\ket{\varphi_d}\ .
\end{subalign}
In Eq.~\eqref{eq:mrow-op-Cnd} we include a factor
\begin{equation}
 \alpha(s_{d}) = \frac{1}{3}\frac{2J+1}{2\times3} \ ,
\end{equation}
where the $1/3$ in the front is due to the antisymmetrization in Eq.~\eqref{eq:Psi-k-sd}, and
the rest covers the spin weights for each individual contribution to the
correlation function.
The factors $2$ and $3$ in the denominator account for the spin $1/2$ of the neutron and the
spin $1$ of the deuteron, respectively.

\subsubsection{Diagrammatic approach for proton-deuteron scattering}
\label{sec:Pionless-pd}

Coulomb effects in Pionless EFT were first studied in Ref.~\cite{Kong:1999sf}
for two nucleons, and in Ref.~\cite{Rupak:2001ci} for proton-deuteron scattering
in the $J=3/2$ channel; Ref.~\cite{Konig:2011yq} was the first to extend this
work to scattering in the $J=1/2$ channel.
Importantly, Ref.~\cite{Vanasse:2014kxa} established that with a nonperturbative
inclusion of Coulomb effects, which is mandatory in the very-low-energy regime,
an isospin-breaking correction to LO three-nucleon force enters at
next-to-leading order (NLO) in the EFT power counting.
At intermediate energies as well as for the trinucleon bound states
(\isotope[3]{H} and \isotope[3]{He}), however, Coulomb effects are a
perturbative correction~\cite{Konig:2015aka,Kirscher:2015zoa,Konig:2016iny}.
In this work we use the nonperturbative treatment in order to describe $pd$
scattering from zero energy all the way up to the breakdown scale of the theory,
$\Mpi\sim140~\MeV$, in a single unified formulation.

The part of the Pionless EFT Lagrangian that is relevant for the present work
can be written as
\begin{multline}
 \mathcal{L} = N^\dagger\left(\ii D_0+\frac{\vD^2}{2\MN}\right)N
 - d^{i\dagger}\left[\sigma_d+\left(\ii D_0+\frac{\vD^2}{4\MN}\right)\right]d^i
 - t^{A\dagger}\left[\sigma_t+\left(\ii D_0+\frac{\vD^2}{4\MN}\right)\right]t^A
 \\
 \null + y_d\left[d^{i\dagger}\left(N^T P^i_d N\right)+\hc\right]
 + y_t\left[t^{A\dagger}\left(N^T P^A_t N\right)+\hc\right]
 + \mathcal{L}_3 + \mathcal{L}_\text{photon} \ ,
 \label{eq:L-Nd}
\end{multline}
with the nucleon field $N$ (with mass $\MN$), a doublet in spin and isospin
space, and two dibaryon fields $d^i$ (with spin 1 and isospin 0) and $t^A$ (with
spin 0 and isospin 1), corresponding to the deuteron and the spin-singlet
isospin-triplet virtual bound state in $S$-wave nucleon-nucleon scattering.
Projectors $P^i_d$ and $P^A_t$ are used to select the appropriate quantum
numbers for nucleon field bilinears.
The formulation in terms of dibaryon fields that we use here is particularly
convenient to discuss nucleon-deuteron scattering.
It is equivalent to Pionless EFT constructed with only nucleon fields in the
strong sector, and the coupling constants $y_{d/t}$ and $\sigma_{d/t}$ can be
related to the standard low-energy constants $C_{0,d/t}$, $C_{2,d/t}$
that multiply two-nucleon contact interactions.
An important feature that arises from the closeness of the low-energy
few-nucleon regime to the so-called unitarity (infinite $S$-wave scattering
lengths) limit---close enough, in fact, to permit a perturbative expansion
around it~\cite{Konig:2016utl}---is the presence of a three-nucleon interaction
already at leading order (LO) in the theory, first derived in
Refs.~\cite{Bedaque:1998km,Bedaque:1999ve}.
We write this interaction in Eq.~\eqref{eq:L-Nd} simply as $\mathcal{L}_3$ and
refer to the review~\cite{Hammer:2019poc} and the original references above for
details.

The coupling of nucleons (and dibaryons) to the electromagnetic (e.m.) field is
implemented by the covariant derivative $D_\mu = \partial_\mu + \ii eA_\mu
\hat{Q}$ with the charge operator $\hat{Q}$, $e^2 = 4\pi\alpha$ the e.m.\
coupling strength, and the photon field $A_\mu$.
The photon kinetic term is included in $\mathcal{L}_\text{photon}$.
In the nonrelativistic low-energy regime we consider, we need only to keep the
contribution of so-called Coulomb photons, corresponding to a static potential
$\sim 4\pi\alpha/(\vq^2+\lambda^2)$ between nucleons, where $\vq$ denotes the
momentum transfer and $\lambda$ is a small photon mass (infrared regulator)
necessary for a momentum-space formulation of the theory.
More details on the formalism can be found in previous publications on the
subject (see \eg.~Ref.~\cite{Konig:2011yq}).

\begin{figure}[tbhp]
\centering
\includegraphics[clip]{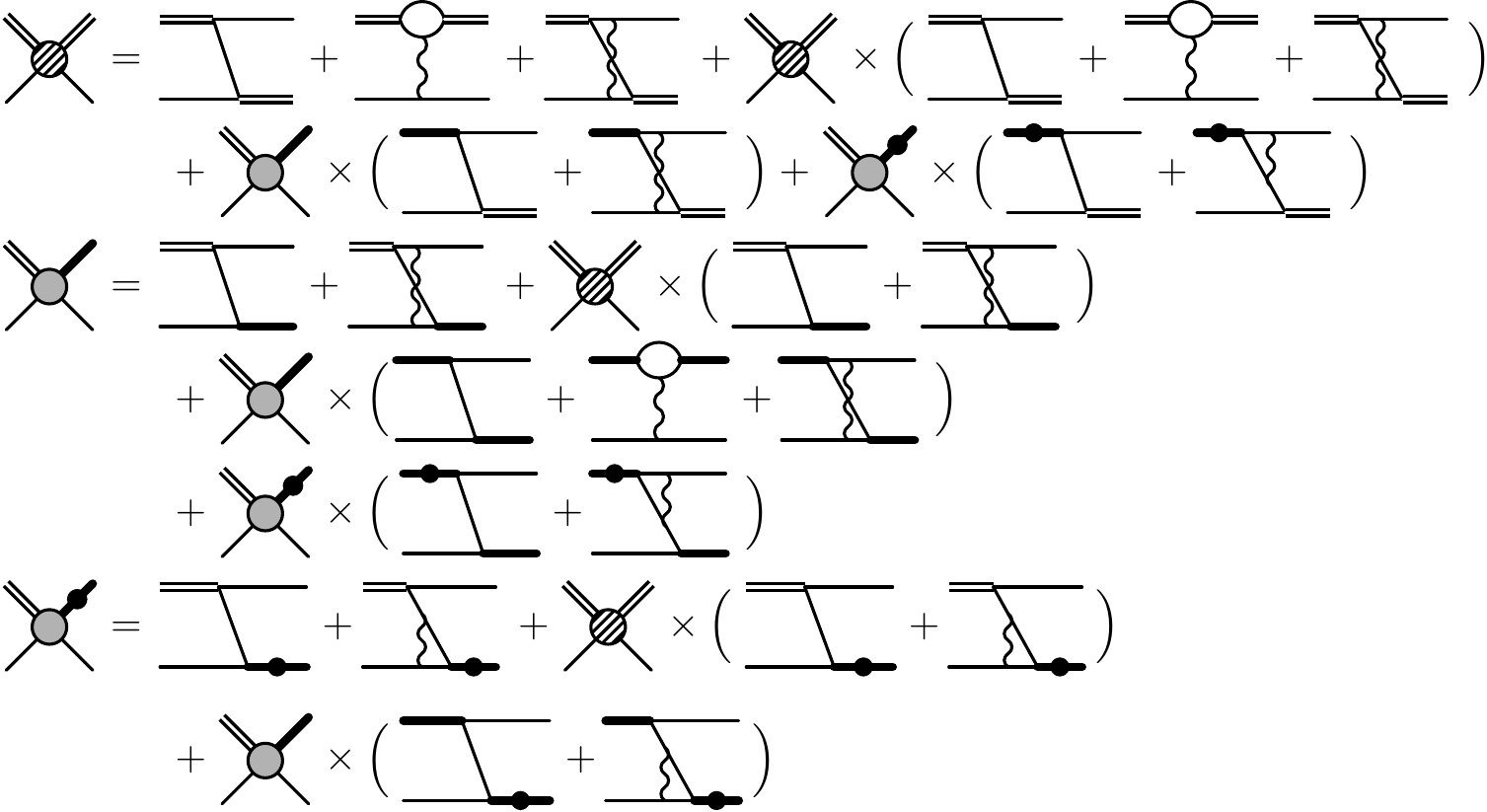}
\caption{Coupled-channel integral equation for the full (\ie, strong + Coulomb)
proton-deuteron scattering amplitude in the $J=1/2$ channel.
The diagrams representing the three-nucleon force have been omitted.
Notation as in Ref.~\cite{Konig:2011yq}.}
\label{fig:pd-IntEq}
\end{figure}

Proton-deuteron scattering in the spin-doublet channels ($J=1/2$) is described by an
integral equation that is shown diagrammatically in Fig.~\ref{fig:pd-IntEq}.
This equation describes an amplitude $\mathcal{T}$, represented by the blob with
hatched shading, and involves two coupled channels because in intermediate
configurations both spin-triplet and spin-singlet two-nucleon states can appear.
These are drawn as double lines and thick lines, respectively.
In spin-quartet channels ($J=3/2$), the Pauli principle prohibits
intermediate spin-singlet states, and consequently in this channel the
scattering amplitude is given by merely the first row in
Fig.~\ref{fig:pd-IntEq}.
For either channel, we numerically implement the integral equation by projecting
on a particular spin channel (described by quantum numbers $s_{d}$ as introduced
in the previous section), and by discretizing all momentum integrals (arising from
loops in the diagrams) to obtain matrix-vector equations.
The full details of this procedure can be found for example in 
Refs.~\cite{Vanasse:2014kxa,Konig:2014ufa}.

We can relate $\mathcal{T}$ to the scattering amplitude $\tilde{T}$
introduced in Sec.~\ref{sec:Faddeev}.
If we consider the special case of a separable two-body interaction between
nucleons, $V(u,u') = C_0 g(u) g(u')$ for momenta $u$ and $u'$ (where $g(u)$ in
an EFT context implements an ultraviolet cutoff for a given regularization
scheme), then we can write (neglecting discrete quantum numbers for simplicity):
\begin{equation}
 \tilde{T}(u_1,u_2)
 = g(u_1) \tau(\MN E-\tfrac34 u_2^2) G_0(E;u_1,u_2) \tilde{\mathcal{T}}(u_2) \,,
\label{eq:T-gG0T}
\end{equation}
where $\tau$ expresses the energy dependence of the separable two-nucleon T-matrix,
\begin{equation}
 t(E;u,u') = g(u) \tau(\MN E) g(u') \,.
\end{equation}
For each two-nucleon channel, the T-matrix can be obtained by algebraically solving
the Lippmann-Schwinger equation for the separable potential $V$~\cite{Konig:2019xxk}, or
equivalently by solving an equation that follows from a diagrammatic representation of
the ``dibaryon propagators'' that appear as intermediate states (double and thick
lines) in  Fig.~\ref{fig:pd-IntEq}~\cite{Vanasse:2014kxa,Konig:2014ufa}.

The relationship between $\tilde{\mathcal{T}}$ and $\mathcal{T}$ is then just a
factor,
\begin{equation}
 \tilde{\mathcal{T}} = {-}\frac{\MN}{4\pi} \mathcal{T} \,,
\label{eq:T-T}
\end{equation}
up to potentially different regularization schemes.
Specifically, the diagrammatic approach does in fact not use the separable Gaussian
regular, but instead imposes a sharp cutoff $\Lambda$ imposed directly on momentum
integrals.
This can be interpreted as setting $g(u) = \Theta(\Lambda - u)$, where $\Theta$ denotes
the unit step function.
Note that the details of the running coupling $C_0(\Lambda)$ change with the regulator,
but the physics does not depend on this arbitrary choice.
Finally, Eqs.~\eqref{eq:T-T} and~\eqref{eq:T-gG0T} can be combined and inserted into
Eq.~\eqref{eq:psi-T}, which then leads to an expression for the $pd$ correlation
function $C_{pd}(k)$ via Eqs.~\eqref{eqs:mrow-op}.
To conclude this part we note that when the momentum $k$ and therefore the associated
energy $E$ is large enough to break up the deuteron ($k \gtrsim 50$ MeV), the Green's
function $G_0$ in Eq.~\eqref{eq:T-gG0T} exhibits an on-shell singularity.
We account for this effect by adding a small imaginary part $\ii\eta$ with $\eta\ll E$
to the energy.

\subsubsection{Next-to-leading order calculation}

In a rigorously perturbative setup, as we employ it here, the $pd$ correlation function
has an expansion of the form
\begin{equation}
 C_{pd}(k) = C_{pd}^{(0)}(k) + C_{pd}^{(1)}(k) + \cdots\ ,
\label{eq:C-pd-expansion}
\end{equation}
where $C_{pd}^{(0)}(k)$ is the leading-order (LO) result, $C_{pd}^{(1)}(k)$ is the
next-to-leading-order (NLO) correction, and the ellipses represent higher-order corrections
that we do not consider in this work.
We emphasize that the LO calculation is still performed in a nonperturbative manner at the
three-nucleon level, which is necessary at least in the $J^\pi = 1/2^+$ partial wave to
generate the \isotope[3]{He} bound state, and for convenience applied to all partial waves at
leading order.
Moreover, the formalism naturally includes the LO two-nucleon interaction 
nonperturbatively, as mandated by the shallow S-matrix poles associated with the large
$NN$ scattering lengths.
All corrections applied on top of LO, however, are included via strict distorted-wave
perturbation theory, and that is what we refer to as the ``rigorously perturbative setup''
in this context.

Assuming that the source operator does not have an expansion by itself, the series in
Eq.~\eqref{eq:C-pd-expansion} is generated by the EFT expansion of the scattering wave function,
\begin{equation}
 \ket{\Psi_k} = \ket{\Psi_k^{(0)}} + \ket{\Psi_k^{(1)}} + \cdots \ ,
\end{equation}
which is, in turn, generated by the expansions for $\tilde{\mathcal{T}}$ and $\tau$,
see Eq.~\eqref{eq:T-T}.
Moreover, the deuteron wave function has an analogous expansion
\begin{equation}
\ket{\varphi_d} = \ket{\varphi_d^{(0)}} + \ket{\varphi_d^{(1)}} + \cdots\ ,
\end{equation}
which gives rise to an expansion for $A_d$.
Overall, we have
\begin{equation}
 A_d^{(0)}C_{pd}^{(0)}(k) = \bra{\Psi_k^{(0)}}{\hat{S}}\ket{\Psi_k^{(0)}}
\end{equation}
at leading order, whereas at NLO we need to extract $C_{pd}^{(1)}(k)$ from
\begin{equation}
 A_d^{(0)}C_{pd}^{(1)}(k) + A_d^{(1)}C_{pd}^{(0)}(k)
 = 2\text{Re}\bra{\Psi_k^{(0)}}{\hat{S}}\ket{\Psi_k^{(0)}} \ .
\label{eq:C-pd-NLO}
\end{equation}
This can be achieved by noting that independently we have $A_d^{(1)} =
2\text{Re}\bra{\varphi_d^{(0)}}{\hat{S}}\ket{\varphi_d^{(1)}}$, and from the LO
calculation we already know $C_{pd}^{(0)}(k)$.
Hence, we can solve Eq.~\eqref{eq:C-pd-NLO} for $C_{pd}^{(1)}(k)$.
\section{Results and comparison}
\label{sec:Results}

\subsection{Proton-proton correlation function}

\begin{figure}[hbtp]
 \centering
 \includegraphics[width=0.65\textwidth]{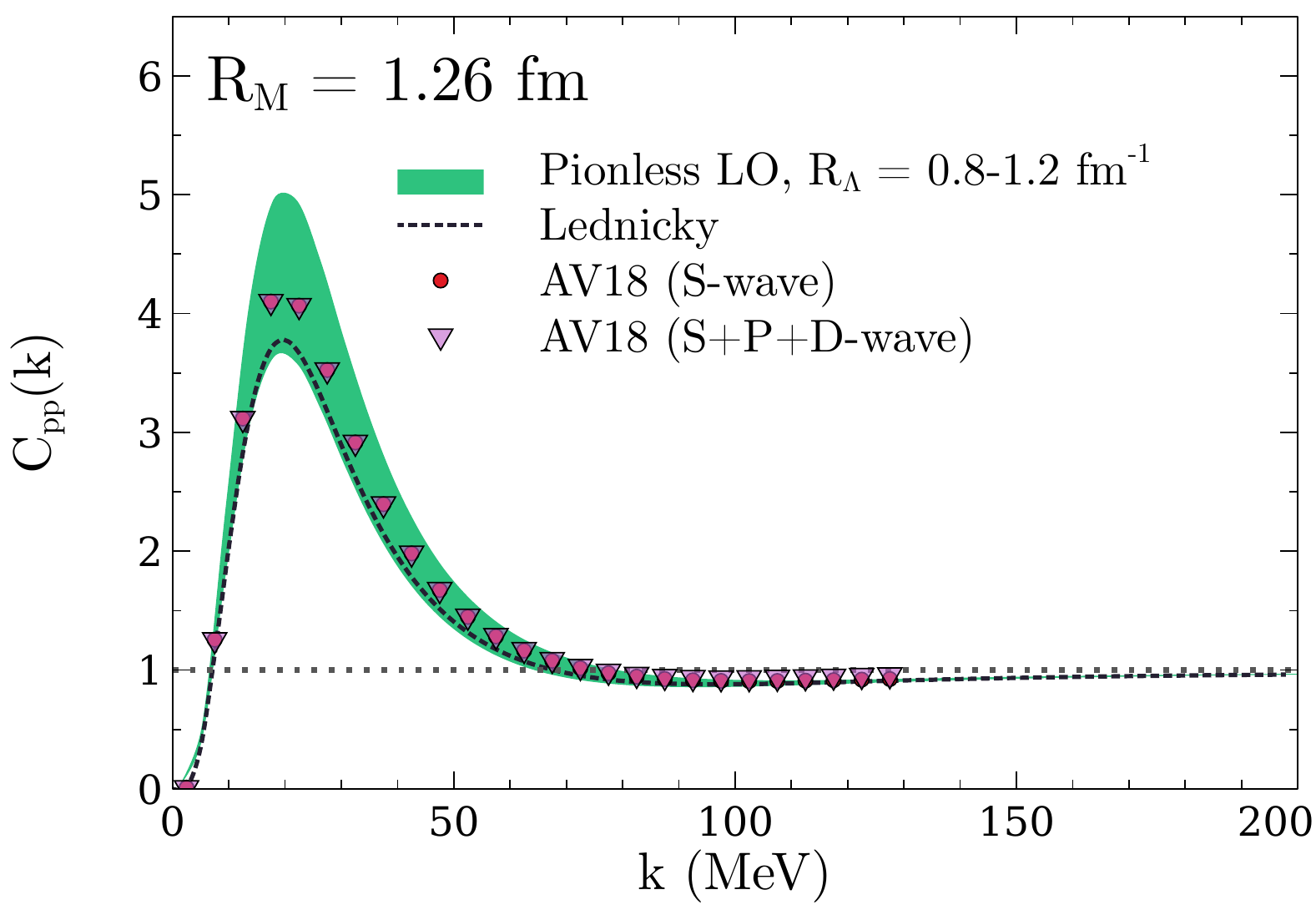}
\caption{%
The proton-proton correlation function $C_{pp}(k)$ as function of the relative momentum $k$ calculated with different approaches.
The symbols show $C_{pp}(k)$ calculated with the AV18 interaction,
with circles representing a calculation that includes the nuclear interaction only in the $S$-wave (${}^1S_0$) channel, while for the triangles the interaction is considered also in the $P$- and $D$-wave channels (the two results are pratically coincident).
The shaded band reflects a Pionless EFT calculation at leading order for
a range of regulator scales (see text for details).
The dashed line shows for comparison a calculation based on the Lednick{\'y} model, as explained in the main text.
\label{fig:Cf-pp}
}
\end{figure}

Before we consider the nucleon-deuteron correlation function, it is instructive
to compare the performance of different nuclear interactions for the 
$pp$ case.
In Fig.~\ref{fig:Cf-pp}, we show the $pp$ correlation function $C_{pp}(k)$
calculated with three different approaches.
The circles and triangles represent the calculation based on the same AV18
potential that we use for the $Nd$ calculation.
For the circles, the nuclear interaction is included only in the S-wave (${}^1S_0$) channel, while for the triangles we include the interaction in 
addition in P- and D-wave channels.
In both cases, additional pure Coulomb contributions are included up to a
maximum angular momentum $\ell_{\text{max}} = 20$.
We observe that the correlation function is completely dominated by the
$S$-wave interaction, with only very small contributions from higher partial waves
for momenta above about 70 MeV.

The green band in Fig.~\ref{fig:Cf-pp} shows the result of a Pionless EFT
calculation at leading order.
For this two-body system, the simplest way to implement Pionless EFT is by
employing a local coordinate-space potential
\begin{equation}
 V_{\text{LO},pp}(r)
 = C_0(R_\Lambda) \exp\!\left({-}\frac{r^2}{R_\Lambda^2}\right)
 + \frac{\alpha}{r} \ ,
\end{equation}
where $R_\Lambda$ is the scale for a local Gaussian regulator (which is
roughly related to a momentum cutoff $\Lambda \sim {2}/{R_\Lambda}$).
For each choice of $R_\Lambda$, renormalization is achieved by adjusting
the coefficient $C_0(R_\Lambda)$ such that $V_{\text{LO},pp}(r)$ overall
reproduces the experimental value for the $pp$ scattering length,
$a_{pp} = {-}7.806$ fm.
The band in the figure is generated by varying $R_\Lambda$ in the range
between 0.8 and 1.2 fm.
We note that for this leading-order calculation only the scattering length
is reproduced exactly, while the next term in the (Coulomb-modified)
effective range expansion, namely the $pp$ effective range, is only
induced by the regulator.
The center of the band shown in Fig.~\ref{fig:Cf-pp} roughly corresponds to
$R_\Lambda = 1.0$ fm, whereas we find the induced effective range closest to
the actual experimental value near $R_\Lambda = 1.1$ fm.
This observation gives good reason to expect that a proper next-to-leading
order calculation---which would fix a second parameter to reproduce the
effective exactly for any $R_\Lambda$---will narrow the band close to the
AV18 result.

Finally, the dashed line in Fig.~\ref{fig:Cf-pp} also shows a calculation of the $pp$ correlation
function based on the so-called Lednick{\'y} model~\cite{Lednicky:2005tb}.
Specifically, the definition of the scattering wave function described in
Eq.~(89) of Ref.~\cite{Lednicky:2005tb} is considered exclusively for the
$S$-wave ($S = J = 0$) contribution.
The overall antisymmetrization of the wave function and the normalization
of the correlation function follows the same approach that is used for
the AV18 $S$-wave contribution.
Similar to Pionless EFT at LO, the only input parameter to this calculation
is the (Coulomb-modified) $S$-wave scattering length $a_{pp}$.
The predicted correlation function agrees with the  Pionless EFT at LO,
but this model disagrees noticeably with both the AV18 calculations 
(with the caveat that neither the AV18 potential nor the Lednick{\'y} model provide a theoretical uncertainty estimate).

\subsection{Comparing $C_{nd}$ using the PHH and Faddeev techniques}

In this subsection we start our study of the $Nd$ correlation function.
We consider first the $nd$ system in order to avoid complications due to the inclusion of the Coulomb interaction.
To this end, it is convenient to write the wave function as in Eq.~(\ref{eq:full1}), taking into account that now the regular and irregular functions reduce to simple spherical Bessel functions,
and the Coulomb phase-shift is set to zero.
Inserting this wave function in Eq.~(\ref{eq:mrow9}), performing the sum over $m_2$ and $m_1$ and realizing that $\int \dd\Omega\, \Psi_{L'S'J'J_z'}^\dag \Psi_{LSJJ_z}$ clearly is not vanishing only
if $J=J'$ and $J_z=J_z'$, one obtains
\begin{equation}
  C_{nd}(k) = \frac{1}{A_d} \frac{1}{6} 4\pi \sum_{JLS} (2J+1)
  \int \rho^5 \dd\rho d\Omega\; \frac{e^{{-}\rho^2/4R_M^2}}{(4\pi R_M^2)^3} 
  \abs{\Psi_{LSJJ_z}}^2
  \equiv \sum_{JLS}  C_{nd}^{LSJ}(k)\ .
\label{eq:mrow11}
\end{equation}
Each component of the wave function $\Psi_{LSJJ_z}$ gives a separate contribution and we can define
\begin{equation}
  C_{nd}^{J^\pi}(k) = \sum_{LS} C_{nd}^{LSJ}(k) \ ,
\end{equation}
where $\pi=\pm$ indicates the parity and the sum is over all the possible $LS$ combinations for a given $J$ and $\pi$. Then we
compare the contributions from different partial waves ${}^{2S+1}L_J$ to the $nd$ correlation function $C_{nd}(k)$.
We use the AV18 interaction and consider the specific cases of $E=0.3195$ MeV ($k=20$ MeV) and $E=2$ MeV ($k=50$ MeV), with
both the PHH method and Faddeev equations.
The results are shown in Table~\ref{tab:PHH-Fad-nd-AV18}.

\begin{table}[htbp!]
\centering
\def\arraystretch{1.3}
\pgfplotstabletypeset
[
  columns = {
   Jpi, Wave, PHH20, Faddeev20, PHH50, Faddeev50
  },
  columns/Jpi/.style = {
   string type, column name=$J^\pi$, column type={c}
  },
  columns/Wave/.style = {
   string type, column name=Wave,
   column type={>{\centering\arraybackslash}p{3em}|}
  },
  columns/PHH20/.style = {
   fixed, zerofill, precision=5, column name=PHH,
   column type={>{\centering\arraybackslash}p{4em}}
  },
  columns/Faddeev20/.style = {
   fixed, precision=5, zerofill, column name=Faddeev,
   column type={>{\centering\arraybackslash}p{4em}|}
  },
  columns/PHH50/.style = {
   fixed, zerofill, precision=5, column name=PHH,
   column type={>{\centering\arraybackslash}p{4em}}
  },
  columns/Faddeev50/.style = {
   fixed, precision=5, zerofill, column name=Faddeev,
   column type={>{\centering\arraybackslash}p{4em}}
  },
  every head row/.style= {
   before row = {
    \TopRule
    \multicolumn{2}{c}{} 
    & \multicolumn{2}{c|}{$k = 20$ MeV} 
    & \multicolumn{2}{c}{$k = 50$ MeV} \\
    \MidRule
   },
   after row = { \MidDoubleRule \rule{0pt}{1.2em} }
  },
  every row no  2/.style = { before row = { \MidRule }, },
  every row no  4/.style = { before row = { \MidRule }, },
  every row no  7/.style = { before row = { \MidRule }, },
  every row no 10/.style = { before row = { \MidRule }, },
  every last row/.style = { after row = { [3pt] \BottomRule } }
]{
Jpi            Wave               PHH20    Faddeev20  PHH50    Faddeev50
${\frac12}^+$  ${}^2S_{\frac12}$  0.42854  0.427631   0.26003  0.259502
\phantom{---}  ${}^4D_{\frac12}$  0.00004  0.000036   0.00075  0.000722
${\frac12}^-$  ${}^2P_{\frac12}$  0.00262  0.002556   0.01662  0.016348
\phantom{---}  ${}^4P_{\frac12}$  0.01713  0.016545   0.07336  0.070621
${\frac32}^+$  ${}^4S_{\frac32}$  0.00931  0.008827   0.01742  0.016513
\phantom{---}  ${}^2D_{\frac32}$  0.00007  0.000065   0.00234  0.002216
\phantom{---}  ${}^4D_{\frac32}$  0.00004  0.000035   0.00115  0.001098
${\frac32}^-$  ${}^2P_{\frac32}$  0.00514  0.004965   0.03225  0.031459
\phantom{---}  ${}^4P_{\frac32}$  0.03588  0.035055   0.13444  0.131181
\phantom{---}  ${}^4F_{\frac32}$  0.00000  0.000000   0.00008  0.000065
${\frac52}^-$  ${}^4P_{\frac52}$  0.06214  0.059001   0.25767  0.245249
\phantom{---}  ${}^2F_{\frac52}$  0.00000  0.000000   0.00007  0.000066
\phantom{---}  ${}^4F_{\frac52}$  0.00000  0.000000   0.00012  0.000097
}
\caption{%
Contributions from partial waves ${}^{2S+1}L_J$ to the neutron-deuteron
correlation function $C_{nd}(k)$ calculated at $E=0.3195$~MeV ($k=20$ MeV) and $E=2$~MeV ($k=50$ MeV) with
the AV18 potential, using two different methods to perform the calculation.
The source radius $R_M$ here is chosen to be $1.5$~fm.
\label{tab:PHH-Fad-nd-AV18}
}
\end{table}

As it can be seen in the table, there is a overall good agreement between the two
calculations up to minor differences.
The remaining small discrepancies are reflecting differences in the numerical
approaches (such as configuration-space versus momentum-space discretizations and
corresponding truncation schemes), and in part they are likely also due to the fact that
for the Faddeev calculation isospin breaking components within the AV18 are neglected,
\ie, the $np$ and $nn$ interactions are taken to be exactly degenerate.
We also observe that at the energies we consider the largest contributions are brought
by the ${}^2S_{\frac12}$ and ${}^4P_{J}$ waves. 

\subsection{Proton-deuteron correlation function}

In the following we show our results for the proton-deuteron correlation function $C_{pd}(k)$,
starting with the PHH calculation.
As in the $nd$ case,  each component of the wave function gives a separate contribution, namely
\begin{equation}
  C_{pd}(k) =  \frac{1}{A_d} \frac{1}{6} 4\pi \sum_{JLS} (2J+1)
 \int \rho^5 \dd\rho d\Omega\;
 \frac{ e^{{-}\rho^2/4R_M^2}}{(4\pi R_M^2)^3} \abs{\Psi_{LSJJ_z}}^2
 \equiv \sum_{JLS}  C_{pd}^{LSJ}(k)\ . 
\label{eq:mrow10}
\end{equation}
As before, we can define
\begin{equation}
 C_{pd}^{J^\pi}(k) = \sum_{LS} C_{pd}^{LSJ}(k) \ .
\end{equation} 

\begin{figure}[hbtp]
\centering
\includegraphics[width=0.65\textwidth]{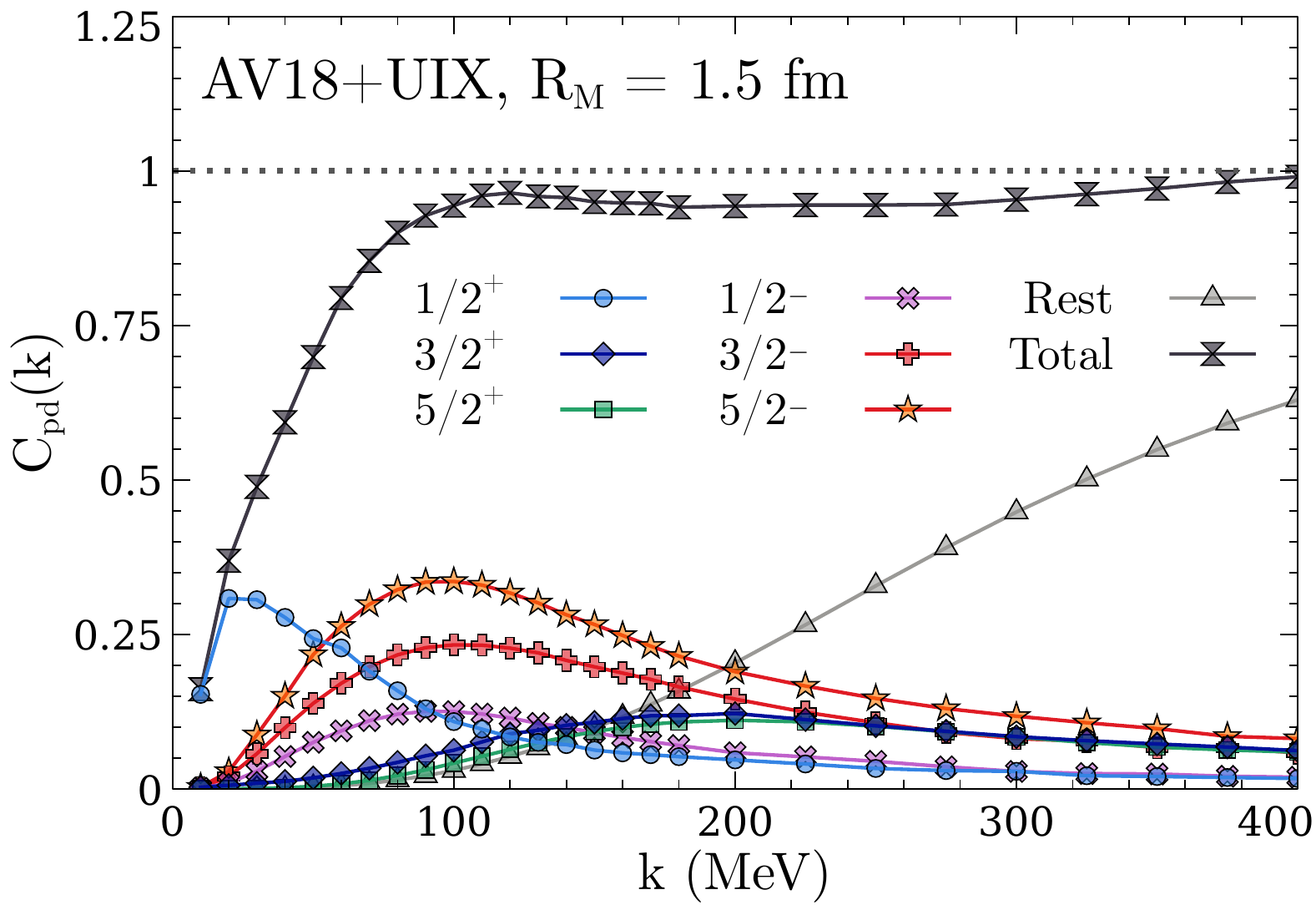}
\caption{The proton-deuteron correlation function $C_{pd}(k)$ and its various contributions $C^{J^\pi}_{pd}(k)$ calculated using the AV18+UIX interaction using the PHH method. The curve labeled ``Rest'' shows the contribution of $J\ge 7/2$ states (they are taken into account via the $\Psi_{m_2,m_1}^{\text{free}}$ of Eq.~(\ref{eq:fullJ})). The calculations are performed using a source size of $R_M=1.5$ fm.}
\label{fig:contJpi3}
\end{figure}

In Fig.~\ref{fig:contJpi3} the $pd$ correlation function $C_{pd}(k)$, calculated using
the AV18+UIX interaction, is shown splitted in the different contribution up to $J=5/2$. For higher values of the angular momentum, the interaction gives a negligible contribution
and the correlation function is therefore computed considering only the Coulomb force.
In the figure this is indicated by the curve labeled ``Rest'', whereas the curve labeled ``TOT'' gives the correlation
function including all contributions. 

Note that low values of $k$ corresponds to small values of the $pd$ relative
kinetic energy $T_{pd}$ (as an example, $k=10$ MeV corresponds to 
$T_{pd}=79$ keV).
For $k\rightarrow0$, the effect of the Coulomb repulsion dominates and
the correlation function tends rapidly to zero.
In this region, the largest contribution is given by the $pd$ waves
with $L=0$, in particular the $L=0$, $S=J=1/2$ wave, whereas the $L=0$, $S=J=3/2$ wave is suppressed at short inter-particle distances due to
the Pauli principle (for $S=3/2$, all three nucleon spins may be aligned). 
Around $k=60$ to $160$ MeV, the $L=1$, $S=3/2$ components with total
angular momentum and parity $J^\pi=1/2^-$, $3/2^-$, and $5/2^-$ start to
give sizeable, resonance-like, contributions.
In fact, in those waves the effective $pd$ interaction is rather attractive
and the corresponding phase shifts increase very fast with energy~\cite{kievsky:2001pdphases}.
Moreover, below $k=200$ MeV there is a moderate splitting of the
quartet $L=1$ phases~\cite{Kievsky:2001fq}, and their relative contributions are nearly related by a factor $(2J+1)$, see Eq.~(\ref{eq:mrow10}).
The effect from this in the total correlation function is the appearance of a wide
bump with maximum located approximately at $k=120$ MeV.
At higher values of $k$, higher partial waves start to contribute and
the correlation function tends to one.

In addition to the study of the different partial waves contribution, it is of interest to consider the correlation function calculated with different interaction models.
To this aim, in Table~\ref{tab:int2} we show the values of the correlation function at several energies computed with the AV18+UIX, the AV18 (without accompanying 3N interaction), and with the NVIa/3N interaction.
Moreover, in the fourth column (labeled by ``Ratio''), the ratio between the second and third columns is given. 
In this way, the effect of the three-nucleon force can be estimated, yielding that for the models considered here it is around $3\%$ in the
region close to $k=100\,$MeV.  

\begin{table}[bth]
\centering
\caption{\label{tab:int2}
The proton-deuteron correlation function $C_{pd}(k)$ calculated
with different interaction models, using the PHH  method.
The source radius $R_M$ is chosen to be $1.5$ fm.
In the fourth column we report the ratio AV18+UIX results (second column) and the AV18 values (third column).}

\def\arraystretch{1.3}
\pgfplotstabletypeset
[
  columns = {
   k, AV18+UIX, AV18, Ratio, NVIa+3N
  },
  columns/k/.style = {
   fixed, zerofill, precision=0, column name=$k$ [MeV],
   column type={c|}
  },
  columns/AV18+UIX/.style = {
   fixed, zerofill, precision=4, column name=AV18+UIX,
   column type={>{\centering\arraybackslash}p{4.5em}}
  },
  columns/AV18/.style = {
   fixed, zerofill, precision=4, column name=AV18,
   column type={>{\centering\arraybackslash}p{4.5em}}
  },
  columns/Ratio/.style = {
   fixed, zerofill, precision=4, column name=Ratio,
   column type={>{\centering\arraybackslash}p{4em}}
  },
  columns/NVIa+3N/.style = {
   fixed, zerofill, precision=4, column name=NVIa+3N,
   column type={>{\centering\arraybackslash}p{4.5em}}
  },
  every head row/.style= {
   before row = { \TopRule },
   after row = {\MidDoubleRule\rule{0pt}{1.2em}}
  },
  every last row/.style = { after row = { [3pt] \BottomRule } }
]{
 k   AV18+UIX  AV18     Ratio    NVIa+3N
 10  0.16132  0.17270  0.93411  0.16097 
 20  0.36898  0.39281  0.93933  0.36899 
 30  0.48880  0.50572  0.96654  0.48863 
 40  0.59314  0.60266  0.98420  0.60124 
 50  0.69867  0.70538  0.99049  0.70721 
 60  0.79433  0.78017  1.01815  0.78195 
 70  0.85442  0.83647  1.02146  0.85364 
 80  0.90001  0.87916  1.02372  0.89545 
 90  0.92780  0.90848  1.02127  0.92104 
100  0.94378  0.92722  1.01786  0.93883 
110  0.96003  0.93669  1.02492  0.95455 
120  0.96443  0.94055  1.02539  0.96181 
130  0.95874  0.93981  1.02014  0.95585 
140  0.95676  0.93746  1.02059  0.94914 
150  0.94977  0.93708  1.01354  0.94305 
160  0.94815  0.93358  1.01561  0.94006 
170  0.94745  0.93038  1.01835  0.93692 
180  0.94122  0.92743  1.01487  0.93683 
200  0.94308  0.92770  1.01658  0.93696 
225  0.94470  0.93082  1.01491  0.93609 
250  0.94476  0.93683  1.00846  0.94315 
275  0.94561  0.94267  1.00312  0.94369 
300  0.95355  0.94905  1.00474  0.94834 
325  0.96230  0.95450  1.00817  0.95536 
350  0.97142  0.96049  1.01138  0.96311 
375  0.98204  0.96682  1.01574  0.97291 
400  0.99051  0.97450  1.01643  0.98617 
}
\end{table}

\begin{figure*}[hbt]
  \centering
 \includegraphics[width=0.65\textwidth]{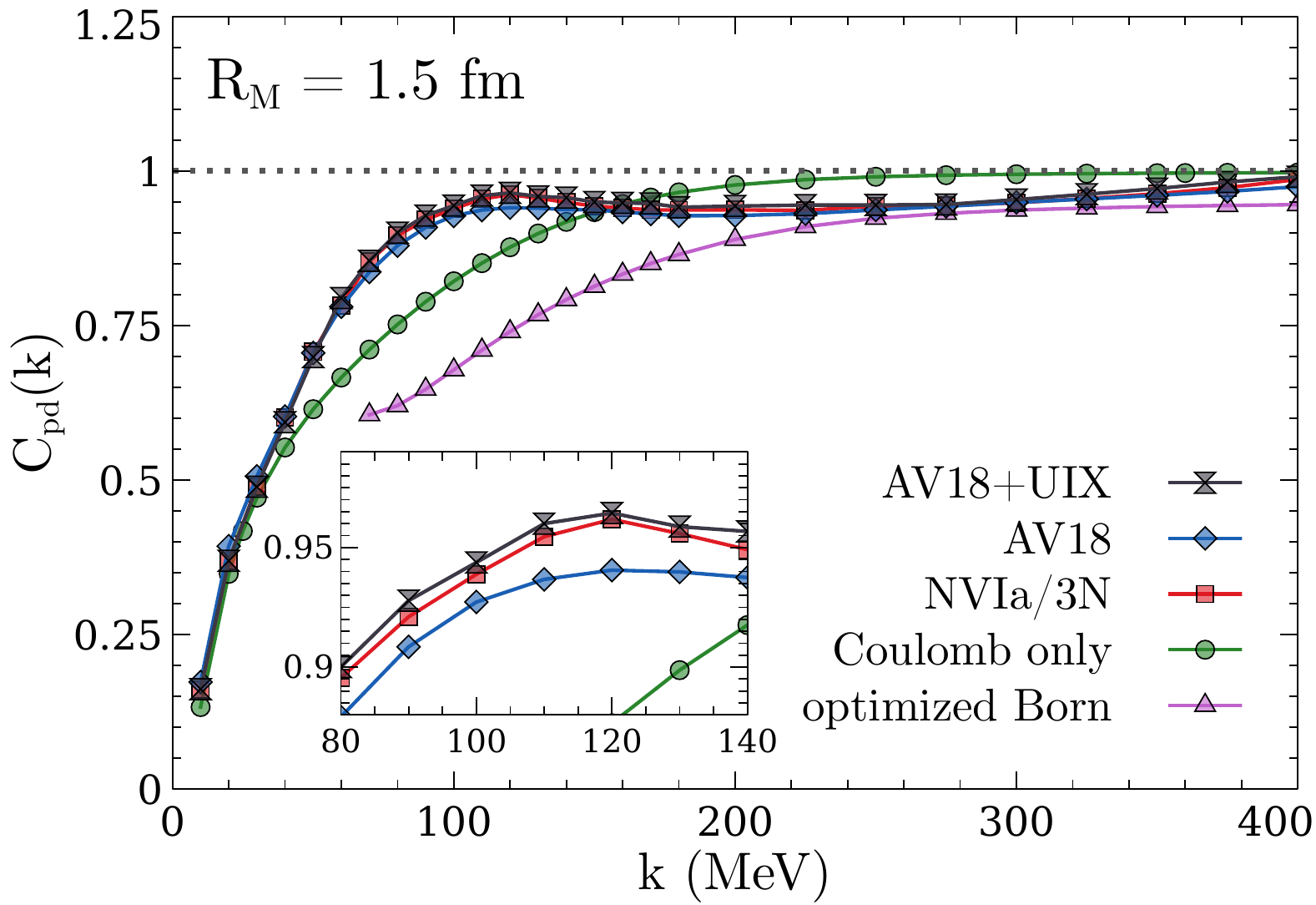}
\caption{The proton-deuteron correlation function  $C_{pd}(k)$ 
calculated with different interactions and approximation of the wave function, using the PHH method. The calculations are performed using $R_M=1.5$ fm. See the main text for more details.
}
 \label{fig:contJpi4}
\end{figure*}

To complete the analysis, in Fig.~\ref{fig:contJpi4} we show the
correlation function calculated with the AV18, AV18+UIX, and NVIa+3N 
interactions, and further comparison calculations that consider only the
Coulomb force and an approach based on the reduction of the wave function 
in the Born approximation. 
The latter two contributions correspond to the following approximations. 
The ``Coulomb only'' curve was obtained considering the free (\ie, pure
Coulomb) $pd$ relative wave function---namely that given in 
Eq.~(\ref{eq:free})---with the deuteron wave function still calculated
with the AV18 interaction.
The difference between this curve and the one labeled ``AV18'' (blue
diamonds) shows the importance of the inclusion of the nuclear 
interaction between the two clusters.
The ``optimized Born'' curve was obtained by neglecting the first term in
Eq.~(\ref{eq:fullJ}), or equivalently setting to zero all the hyperradial 
functions $u_{n,\alpha}(\rho)$.
In this case, the wave function is approximated by the asymptotic terms 
given in the second, third and fourth lines of Eq.~(\ref{eq:fullJ}).
Then the T-matrix elements $T^J_{LS,L'S'}$ are determined from the Kohn
variational principle, using that wave function as the trial input.
This approximation works better for high partial waves in which the
centrifugal barrier suppresses the effects of the interaction~\cite{kievsky:1996ba}.
In fact, for $S$ and $P$ waves, this approximation gives rather different 
results from those obtained using the full wave function.
Therefore, the difference between the curves obtained with the full wave
function and that labeled ``optimized Born'' shows the importance of the
``distortion'' of the deuteron in the process. 
In other word, the $Nd$ wave function at short distances is not simply
given by the product of the deuteron wave function times the spin state
of the third particle, but a full treatment of the three-body dynamics is 
necessary.
Note that for $k<60$ MeV, this ``optimized Born'' approximation
predicts a completely wrong correlation function, which therefore has not 
been reported in the figure.

\begin{figure}[hbtp]
 \centering
 \includegraphics[width=0.65\textwidth]{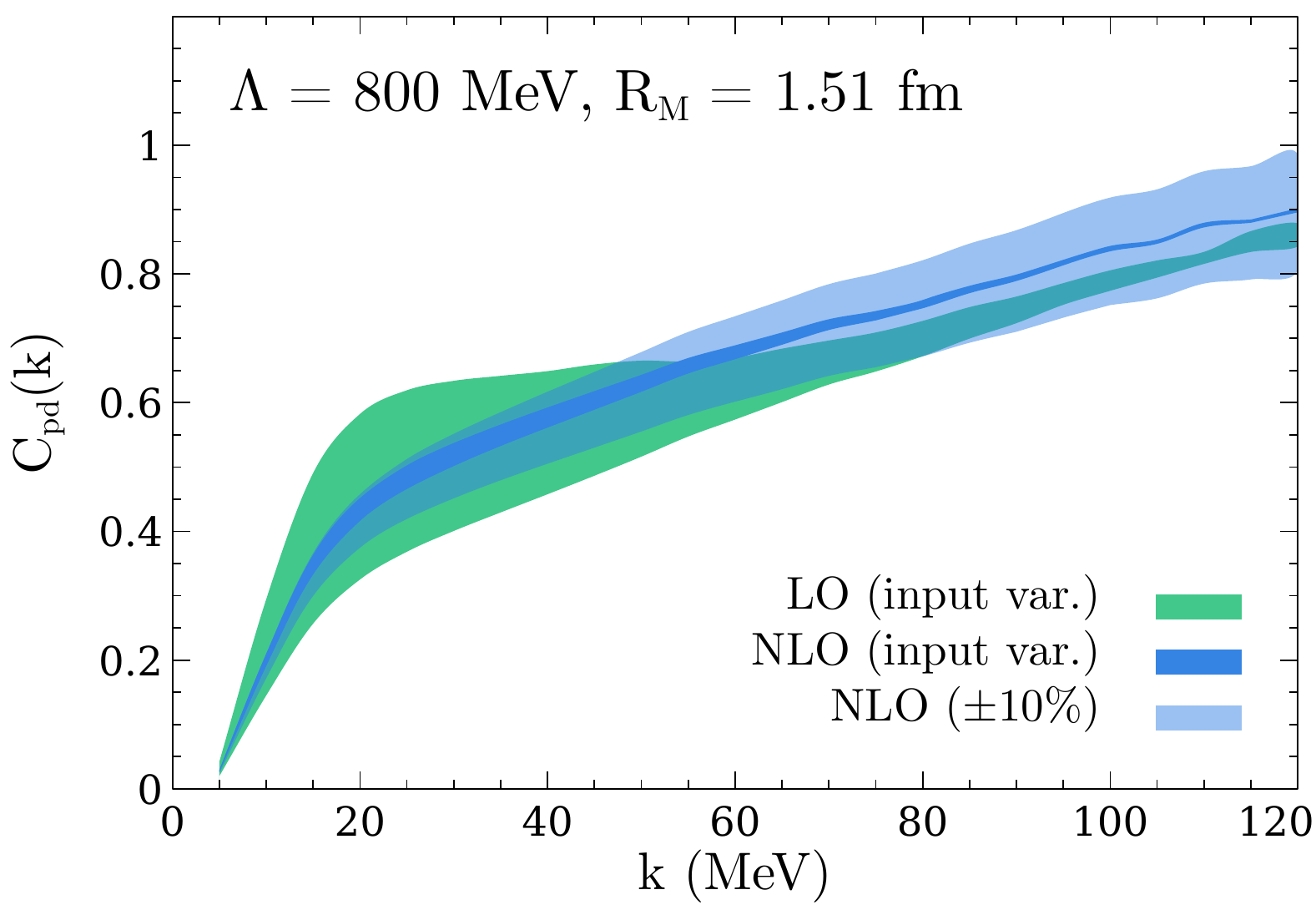}
\caption{%
Proton-deuteron correlation function calculated in Pionless EFT for a source radius
$R_M=1.51$ fm.
The shaded bands here represent the theoretical uncertainty from the EFT expansion.
At LO, this is estimated by varying the EFT input parameters for both the ${}^3S_1$
two-body interaction as well as for the three-nucleon force (see text for details).
At NLO, technical restriction at present permit us to only vary the input for the
three-nucleon force, while the ${}^3S_1$ two-nucleon remains fixed to reproduce the
deuteron at its physical binding energy.
As an additional lighter band we therefore include a blanket 10\% variation to show
a crude \textit{a priori} estimate of the NLO uncertainty.
\label{fig:Cf-pd-var-LO-NLO}
}
\end{figure}

The main result of the Pionless EFT calculation for $C_{pd}(k)$ is summarized in
Fig.~\ref{fig:Cf-pd-var-LO-NLO}.
For a fixed source radius $R_M = 1.51$ fm this figure shows the correlation function
at LO and NLO in the EFT expansion as shaded bands, reflecting the theoretical
uncertainty stemming from the EFT expansion.
This calculation explicitly includes the nuclear interaction in $pd$ $S$- and $P$-waves
(which are all degenerate with respect to the total spin $J$ at this order) and
adds pure Coulomb (or Bessel, in the $nd$ case) contributions on top of the interacting
waves up to total angular momentum $\ell_{\text{max}} = 15$.

To generate the uncertainty estimate, we have varied the input parameters that
enter in the EFT renormalization conditions.
Specifically, the contributions that primarily affect the calculation at LO are 
the ${}^3S_1$ two-nucleon interaction and the three-nucleon contact interaction.
The former can be determined by reproducing either the exact deuteron binding
energy or the experimental value for the scattering length in that channel.
At LO, these choices are equivalent from the EFT perspective.
Fitting $C_0(\Lambda)$ to reproduce the ${}^3S_1$ scattering length yields a deuteron
underbound at about $1.4$ MeV, which gets moved close to the experimental binding
energy by perturbative NLO corrections.
Similarly, the three-body interaction can be fit to reproduce either the experimental
triton binding energy or the $nd$ scattering length (either way, the splitting between
the \isotope[3]{H} and \isotope[3]{He} binding energies is a prediction at this
order that has been studied in great detail~\cite{Ando:2010wq,Koenig:2011lmm,Vanasse:2014kxa,Konig:2014ufa,Konig:2015aka,Kirscher:2015zoa}).
The LO band in Fig.~\ref{fig:Cf-pd-var-LO-NLO} is based on the maximum variation
from overall four different choices of input combinations.
We chose not to vary the ${}^1S_0$ input here because the scattering length in this
channel, $a_{{}^1S_0} = {-}23.7$ fm is so close to the unitarity limit that small
deviations from this value would hardly make any difference.

Conceptual constraints prevent us at present from performing an NLO calculation
with the ${}^3S_1$ input fixed to the scattering length because, as mentioned above,
range correction will shift the deuteron binding energy.
Further theoretical work is required to derive a perturbative nucleon-deuteron
scattering formalism that can handle the moving threshold arising from the
expansion in the two-nucleon sector.
The darker NLO band in Fig.~\ref{fig:Cf-pd-var-LO-NLO} is therefore limited to
varying the input for the three-nucleon interaction, in the same way as described above.
In addition, we show a lighter NLO band generated from a $\pm10\%$ variation
of $C_{pd}(k)$ around the result where the three-nucleon force is fit to reproduce 
the $nd$ scattering length.
We note that although this approach reflects the \textit{a priori} estimate for
the Pionless EFT uncertainty at NLO, it does not take into account the constraint
that the correlation function should approach unity as $k 
\rightarrow\infty$.

Generally, that constraint would be expected to lower the NLO uncertainty.
However, for larger $k$ we observe that $pd$ $P$-wave contributions start dominating
over the $S$-wave, and ultimately higher $pd$ partial waves become sizeable as
well.
It is known that $P$-wave $Nd$ phase shifts converge relatively slowly in Pionless EFT, with sizeable corrections at N2LO.
While that calculation is currently beyond our reach, we expect that it will
improve agreement of the Pionless EFT calculation with the results from potential
models.
Moreover, at LO and NLO, Pionless EFT receives its two-nucleon input from
$S$-waves only, while all higher partial waves vanish by construction at these
orders (keep in mind that the three-nucleon dynamics nevertheless induce $P$ and
higher partial waves in the $Nd$ sector).
These effects will also enter at N2LO and induce a splitting between contributions
from different $J$ for the same $l_2,s_2$ combination~\cite{Vanasse:2013sda}.

It should be noted, however, that the breakdown scale of Pionless EFT is expected
to be set by the pion mass $M_\pi\sim140$ MeV.
One should therefore not expect this EFT to perform well for momenta $k$ near
or beyond that scale; the expansion is constructed for the low-energy regime.

As mentioned at the end of Sec.~\ref{sec:Pionless-pd}, for calculations above
the deuteron breakup threshold we include a small imaginary part $\ii\eta$ in
the energy to regularize an on-shell singularity.
For the results shown in Fig.~\ref{fig:Cf-pd-var-LO-NLO}, we have used a value
$\eta = 0.1$ MeV.
Increasing this to $\eta = 0.5$ MeV leads to a variation of typically about 
$1\%$, which is negligible compared to other uncertainties.
Similarly, we used a regulating photon mass $\lambda = 0.2$ MeV for all
calculations shown here and note that variations due to alternative choices
can be be safely neglected.

\begin{figure}[hbtp]
 \centering
 \includegraphics[width=0.65\textwidth]{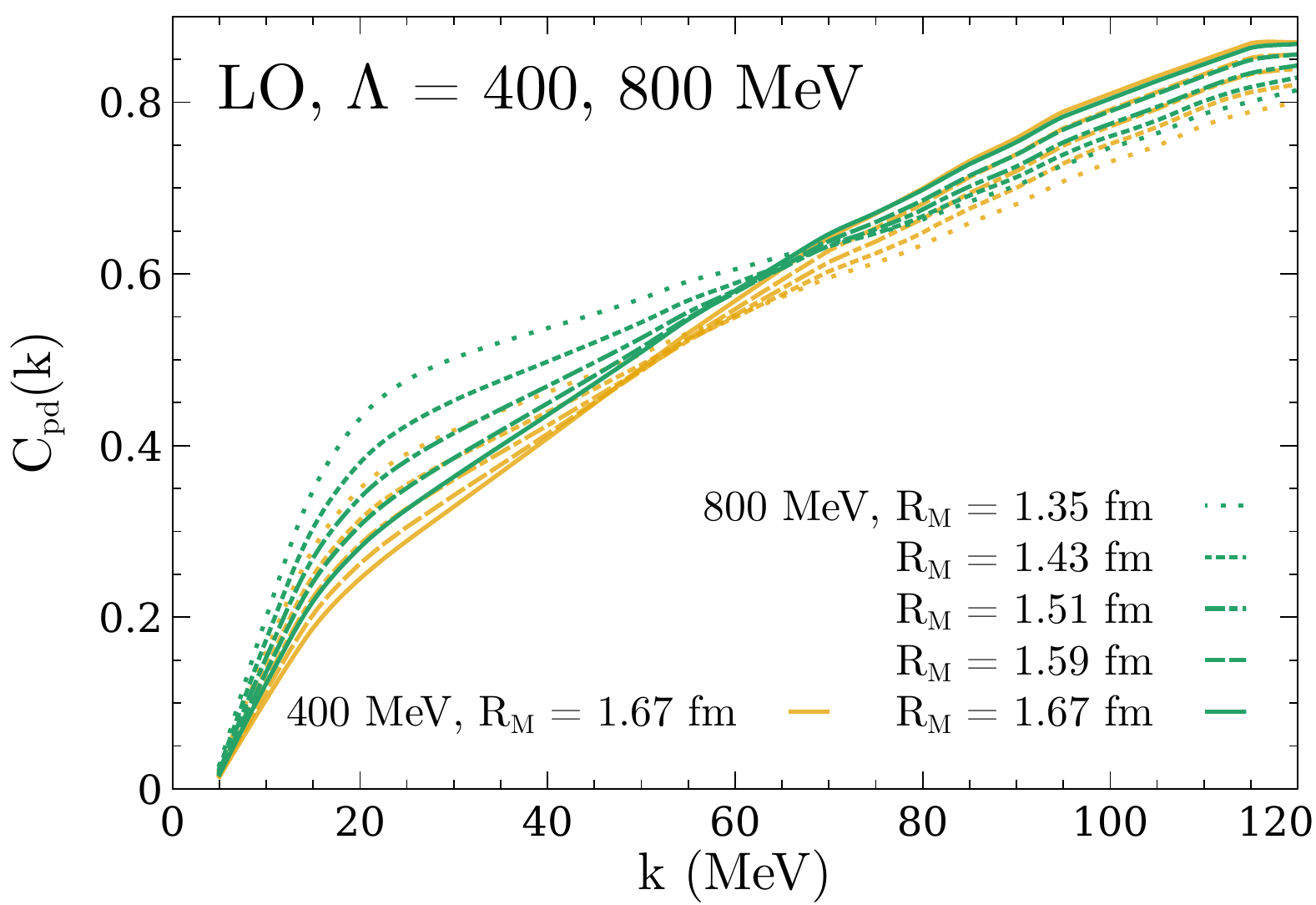}
\caption{%
The proton-deuteron correlation function $C_{pd}(k)$ calculated in Pionless EFT for different
EFT cutoffs $\Lambda$ and source radii $R_M$.
\label{fig:Cf-pd-LO-400-800}
}
\end{figure}

For the results shown in Fig.~\ref{fig:Cf-pd-var-LO-NLO} we used a regulator scale
(cutoff) $\Lambda = 800$ MeV.
In the diagrammatic framework we used, this is implemented with a sharp upper bound
on momentum integrals at the three-nucleon level, while the two-nucleon subsector
is treated using dimensional regularization.
Pionless EFT like any effective field theory exhibits some residual cutoff
dependence that should decrease in magnitude as one goes to higher orders, and
it is an indication of proper renormalization that results for observables
overall flatten out at large cutoffs.
In Fig.~\ref{fig:Cf-pd-LO-400-800} we show the change in the LO $pd$ correlation
function as one goes from $\Lambda = 400$ to $\Lambda = 800$ MeV; little additional
variation is observed for larger $\Lambda$.
In this figure we also use different line styles to show how the correlation function
changes as we vary the source radius $R_M$ between $1.27$ and $1.59$ fm.

\subsection{Proton-deuteron vs.\ neutron-deuteron correlation function}

\begin{figure}[hbtp]
 \centering
 \includegraphics[width=0.65\textwidth]{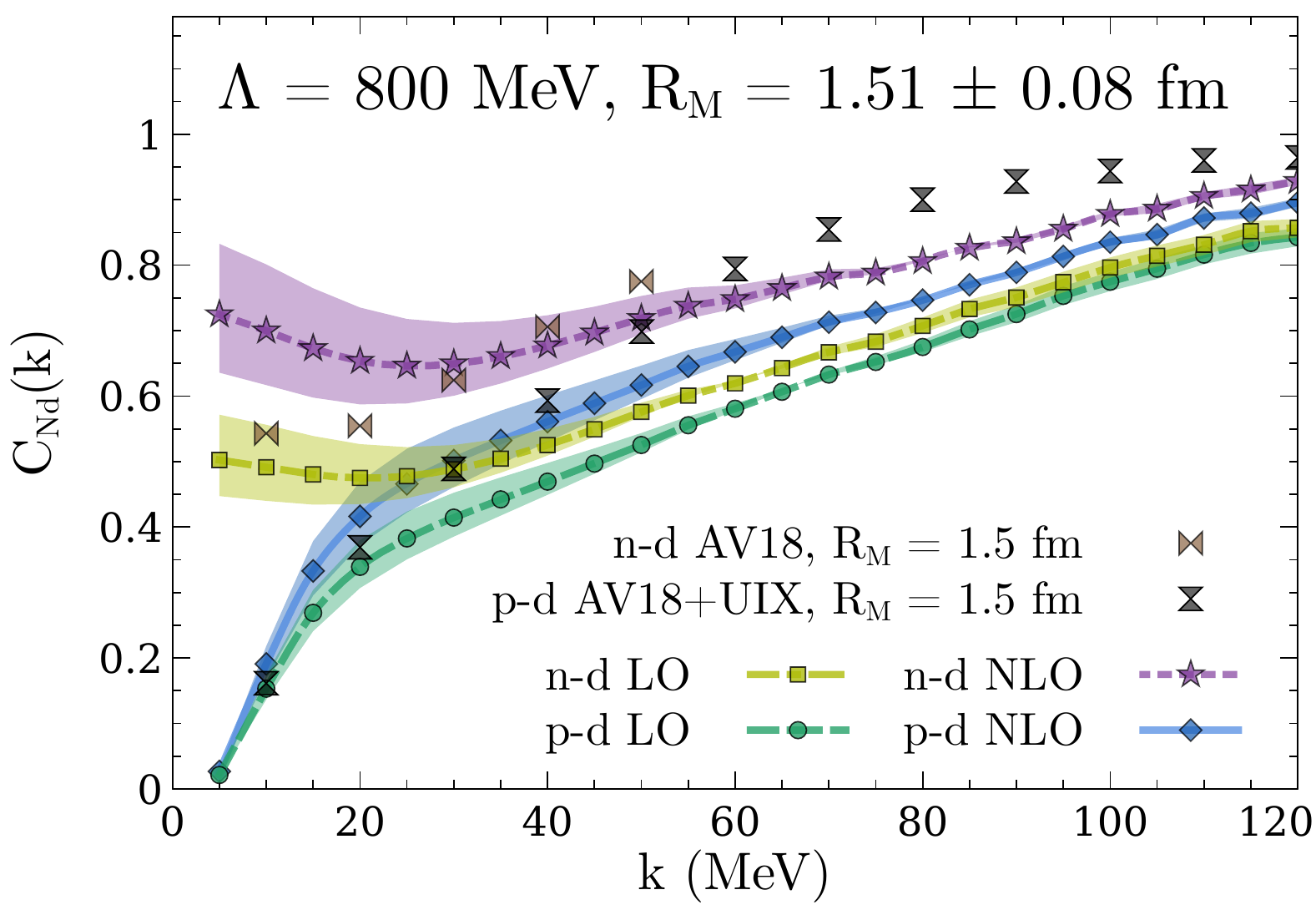}
\caption{%
The proton-deuteron and neutron-deuteron correlation functions calculated in Pionless EFT with $\Lambda=800$ MeV.
The central curves show results for a source radius $R_M = 1.51$ fm, while shaded bands
here indicate the result of varying the source radius by $\pm0.08$ fm.
Double triangle show $nd$ AV18 and $pd$ AV18+UIX results calculated at $R_M = 1.5$ fm for comparison.
\label{fig:Cf-pd-nd-LO-NLO}
}
\end{figure}

In Fig.~\ref{fig:Cf-pd-nd-LO-NLO} we compare the correlation functions for
$pd$ and $nd$ systems.
For this comparison we keep the EFT input paramters fixed (with the ${}^3S_1$
channel fixed to reproduce the deuteron binding energy and the three-nucleon
force fit to the $nd$ scattering length) and show as shaded bands the result
of varying the source radius $R_M$ around a central value of $1.51$ fm.
Consistent with the expectation that Coulomb effects should be a perturbative
effect anywhere except at the lowest energies, we observe that the $pd$ and $nd$ 
curves in Fig.~\ref{fig:Cf-pd-nd-LO-NLO} approach one another with increasing
momentum $k$.

We include in Fig.~\ref{fig:Cf-pd-nd-LO-NLO} also results obtained for the $nd$
correlation function with the AV18 potential (without additional three-nucleon
force, calculated in momentum space via Faddeev equations), as well as for the $pd$ correlation function from an AV18+UIX potential (calculated in coordinate-space via the PHH method).
For these calculations we use a source radius $R_{M} = 1.5$ fm.
Overall these results and the Pionless EFT calculation are in reasonable agreement, in particular if one keeps in mind that in this figure we do not indicate EFT
uncertainty bands on top of the source-radius variation.
For the $nd$ correlation function, some mild tension between the EFT result and potential models \emph{might} exist in the low-$k$ region.
While based on comparing $C_{pd}(k)$ calculated using AV18 with and without UIX
three-nucleon force we do not expect including UIX in the calculation of $C_{nd}(k)$
would improve the agreement, we note that even at small $k$ there is a sizeable
shift from LO to NLO in the EFT result.
Based on that, we believe that an N2LO calculation, which would include $NN$
$P$-wave interactions as well as effects from the ${}^3S_1$-${}^3D_1$ mixing
induced by the nuclear tensor force, is likely to narrow the discrepancy
between the different interactions.
\section{Summary and outlook}
\label{sec:Summary}

Femtoscopic analyses of correlation functions extracted from high-energy
collisions of protons and nuclei have opened the door to new studies of
low-energy scattering processes in light systems such as $pd$, $\Lambda d$,
$ppp$, $pp\Lambda$, and many others.
Measurements of correlations in these systems have recently been 
performed, or are planned by the ALICE Collaboration in the near future.
Accordingly, methods that have been applied in recent years to calculate
scattering observables can be used to obtain the above-mentioned 
correlation functions.
The present study that performs a detailed analysis of the $Nd$ correlation
functions is the first step in this direction.
Although the $nd$ correlation function, $C_{nd}$, cannot be measured at
present since neutron detectors are not being used in the relevant experiments,
its study serves to compare different methods, as the Faddeev and PHH 
techniques.
In fact, the $nd$ system does not present the challenge of treating the
long-range Coulomb interaction.
In this work, the AV18 potential has been used to make comparisons for $C_{nd}(k)$, with the
conclusion that the PHH technique and the solution of the Faddeev equations
produce extremely close results.
This study, which directly involves the scattering wave functions, extends to
some extent previous benchmarks done between these techniques~\cite{Huber:1995zza}.

The next step in this work has been to use the PHH wave functions obtained for
AV18 and other nuclear potential models to compute the $pd$ correlation function
$C_{pd}(k)$ in a broad energy range, in order to enable detailed comparisons to
current and upcoming measurements. The ALICE Collaboration has presented preliminary results for the $pd$ correlation function measured in proton-proton collisions~\cite{Singh:2022qmg} and final results are expected to be published soon.
The correlation function, as a function of the energy, has some structure produced by the
interplay of contributions from different partial waves.
At low energies the system in relative $S$-wave is dominant, whereas a peak around
values of $k=120\,$MeV appears when the relative $P$-wave starts to dominate.
These are the partial waves in which the short-range nuclear interaction produces
the largest effect.
Due to the centrifugal barrier, higher partial waves are mostly dominated by the
Coulomb interaction.
All these considerations have been presented in dedicated figures and
tables.
In particular, we have considered the impact of different interactions,
with and without the inclusion of three-nucleon forces, on the correlation functions.
The conclusion is that within the context of potential models different interactions
give very small variations, not above $1\%$ effects, whereas the three-nucleon force
produces changes of around $2\%$ in the observable.
Since the correlation function is an integral observable, effects of this kind are expected to be small. However a $2\%$ effect is likely within the reach of the next
experimental runs planned by the ALICE Collaboration, and this is one of the main
indications of the present analysis.
In addition to phenomenological potential models, we have also performed a Pionless
EFT calculation of the correlation function, going up to next-to-leading order in
the EFT expansion in a rigorously perturbative setup.
Within the theoretical uncertainty of the EFT, we find overall good agreement with
the potential-model calculations in the low-energy regime where the EFT is
applicable.

In order to compute the correlation function, two ingredients are needed, the source function and the scattering wave function.
The size of the source is determined by the size of the emitting nucleon source and it is fixed by the analysis of the transverse mass 
$m_T$ (defined as
$\smash{m_T = \left(k_T^{2} + m^{2}\right)^{1/2}}$, where $m$ is the average mass and $k_T$
is the transverse momentum of the pair). A precise determination of the dependence of the source size with the transverse mass $m_T$  has been realized in proton-proton collisions~\cite{ALICE:2020ibs}.
For the present analysis, the source term is characterized by the effective nucleon-nucleon distance and depends on the $m_T$ of the emitted $pd$ pairs.
The value of $R_M=1.5$ fm was used in this paper because it is close to the value that gives the best description of the preliminary $C_{pd}(k)$ data from the ALICE Collaboration~\cite{Singh:2022qmg}.

Overall, we can draw two main conclusions: the first is that the nucleon-deuteron
scattering wave function, calculated in the present analysis with a full
account of the three-body dynamics, introduces a complex dynamical
behavior in the correlation function through the relative importance of 
different partial waves, in particular the interplay between $S$- and 
$P$-waves. 
Secondly, we show that over the considered range of momenta up to 400 MeV, the
correlation function is sensitive to aspects of the nuclear interaction, in the
present work constructed as a sum of two- and three-nucleon contributions.
We conclude that the present study supports the experimental efforts devoted to
measuring the correlation function in light nuclear systems dominated by
the strong interaction.

\begin{acknowledgements}
We thank Laura Fabbietti, Johann Haidenbauer, and Stanislaw Mr\'owczy\'nski for useful discussions.
S.K.~acknowledges discussions with participants of the INT Program 
INT-23-1a, ``Intersection of nuclear structure and high‐energy nuclear 
collisions'' thanks the Institute for Nuclear Theory for its hospitality.
This work was supported in part by the National Science Foundation under Grant
No. PHY--2044632.
This material is based upon work supported by the U.S. Department of Energy,
Office of Science, Office of Nuclear Physics, under the FRIB Theory Alliance,
award DE-SC0013617.
Computational resources for parts of this work were provided by the Jülich
Supercomputing Center as well as by the high-performance computing
cluster operated by North Carolina State University.  We also gratefully acknowledge
the support of the INFN-Pisa computing center.

\end{acknowledgements}

\appendix

\bibliography{Content/Bibliography}

\end{document}